# iDART: Interferometric Dual-AC Resonance Tracking for Nano-Electromechanical Mapping

*J. Bemis[1], F. Wunderwald[2], U. Schroeder[2], X. Xu[3], A. Gruverman[3], and R. Proksch,[1,*]*

1 Asylum Research, Oxford Instruments, Santa Barbara, CA, USA

2 Namlab, Dresden, Germany

3 Department of Physics and Astronomy, University of Nebraska, Lincoln, NE 68588, USA

* roger.proksch@oxinst.com

# Abstract

Piezoresponse force microscopy (PFM) has established itself as a very successful and reliable imaging and spectroscopic tool for measuring a wide variety of nanoscale electromechanical functionalities. Quantitative imaging of nanoscale electromechanical phenomena requires high sensitivity while avoiding large bias artifacts. Conventional PFM often relies on relatively high voltages to overcome optical detection noise, leading to various non-ideal effects including electrostatic crosstalk, Joule heating, and tip-induced switching. To mitigate this situation, we introduce interferometrically detected, resonance-enhanced dual AC resonance tracking (iDART), which combines femtometer-scale displacement sensitivity of quadrature phase differential interferometry with contact resonance amplification. Through this combination, iDART achieves at least a tenfold improvement in signal-to-noise ratio (SNR) improvement over current state-of-the-art PFM approaches, including both single-frequency interferometric PFM or conventional, resonance-enhanced PFM using optical beam detection. This allows a >10x improvement of imaging sensitivity. We demonstrate this improvement on PZT and a variety of hafnium-based ferroelectric thin films. Switching spectroscopy shows similar improvements, further revealing reliable hysteresis loops at small biases, mitigating nonlinearities and device failures that can occur at higher excitation voltage amplitudes. These results position iDART as a powerful approach with extremely high SNR suitable for probing piezo- and ferro-electrics. These results establish iDART as a sensitive, low-bias approach for nanoscale electromechanical imaging and spectroscopy of weak piezoelectric systems and extending functional imaging capabilities to 2D ferroelectrics, beyond-CMOS technologies, and biomaterials.

# Introduction

Piezoresponse force microscopy[1] (PFM) has become an indispensable tool for nanoscale electromechanical imaging. Several approaches are commonly used to map piezoresponse. The original and still very common approach is to apply a single frequency, sub-resonant excitation bias to the cantilever and then measure the resulting amplitude and phase.[1] This approach has the benefits of being simple to implement and interpret, where the amplitude $A$, given by $A = d_{eff} V_{ac}$, is usually interpreted as a measure of the localized inverse piezoelectric coefficient $d_{eff}$ multiplied by the voltage amplitude of the AC bias, $A = d_{eff}V_{ac}$ The associated phase of the response, referenced to the phase of the AC bias is associated with the direction of the ferroelectric polarization under the tip.

Recent trends in 2D ferroelectrics and low-power, beyond Moore's law computing[2] involve materials that can have both small electromechanical sensitivities and low breakdown voltages. This provides a double challenge for measurements, requiring increased sensitivity without simply increasing the bias $V_{ac}$.

In addition to breakdown, intense electric fields concentrated at the nanoscale tip–sample junction can drive polarization changes, along with ionic and electrochemical processes.[3] While polarization switching is exploited in Switching Spectroscopy PFM (SSPFM),[8] unintended domain writing during nominally passive imaging can be a major artifact. The domain structures being studied can be inadvertently modified by large biases, undermining reliability and reproducibility and complicating interpretation. One example is shown in supplemental Fig. S3, where domain structures in a PZT sample change under the influence of a large ac bias ($V_{ac} = 7V$).

Examples of ionic motion include migration of oxygen vacancies, [4] charge injection, and local redox reactions. Such processes cause volumetric strain, charge trapping, and irreversible surface modification. These effects are nonlinear with respect to applied bias and are more pronounced at larger voltages. These effects may overwhelm the PFM response or can introduce hysteresis, drift, and history dependence into the measurements, complicating interpretation.[5] When large AC or DC biases are applied, leakage currents through the tip–sample junction can also generate Joule heating that, in turn, may result in localized dimensional changes that can be falsely interpreted as due to piezoresponse. In addition, localized heating may degrade the sample or alter local polarization, leading to spurious or unstable signals or being a precursor to shorting out the sample.

A useful voltage scale for minimizing perturbative PFM is the thermal equivalent voltage, $V_{therm} = k_B T/e \approx 25.7mV$, where $k_B$ is Boltzmann's constant, $T$ is the absolute temperature, and $e$ is the electron charge. Using small AC biases on the order of a few $V_{therm}$ or less helps ensure gentle imaging of materials without the effects discussed above and may in fact be essential for stable imaging of weak materials. Applying a tip bias V is larger than a few $V_{thermal}$ can dramatically increase the likelihood of tip-induced switching, charge injection, electrochemical activity, [6] or ionic migration. [7] By contrast, operating at biases near or below a few $V_{thermal}$ helps ensure that the measurement only probes the linear piezoelectric response without significantly altering the underlying energy landscape. These limitations are particularly acute in materials with intrinsically weak piezoresponse, such as hafnia-based ferroelectrics, two-dimensional ferroelectrics such as $In_2Se_3$, SnSe, and van der Waals heterostructures.[8], [9], [10], [11], [12], [13] Antiferroelectric materials present a related challenge: while they exhibit large field-induced responses at high bias, their small-signal response near equilibrium is nearly zero. [14], [15]

# Part I. Challenges of PFM at Small Bias

PFM probes nanoscale electromechanical coupling by applying an electrical bias through a conductive AFM probe and detecting the induced displacement. It is one of the few techniques capable of simultaneously actuating and sensing ferroelectric domains at nanometer length scales, making it central to fields ranging from nonvolatile memory development to emerging 2D ferroelectrics. While the method is conceptually simple, practical implementation faces a number of intertwined challenges. Achieving high sensitivity while avoiding measurement-induced artifacts requires balancing signal strength against bias-induced distortions and various types of crosstalk.

A fundamental limitation of conventional PFM stems from the need to apply an AC bias large enough to overcome displacement detector noise. The problem is particularly severe for weak ferroelectrics such as hafnia-based thin films, antiferroelectrics, and 2D ferroelectrics, where large biases, in addition to driving piezo- and other electromechanical motion, can induce electrostatics, surface chemistry, Joule heating, and tip-induced switching, thereby distorting or destroying the very states being probed. As discussed above, if the AC bias is less than a few $V_{therm}$, we expect nonintrusive measurements. As the AC bias grows beyond a few $V_{the}$ , we expect an increased likelihood of other, usually undesirable, bias-induced effects.

Small values of $V_{ac}$ place demands on PFM cantilever detection sensitivity and noise. As mentioned above, the first PFM[1] measured the response to a single, sub-contact resonance modulation bias. Neglecting large 1/f noise, sub-resonant motion of commonly used PFM cantilevers (for example, Adama 2.8 conductive diamond probes[16]) being measured with optical beam deflection[17] (OBD) have detection noise floors as low as $S_{SF} \geq 100 - 200\ fm/\sqrt{Hz}$. In most practical applications with a measurement bandwidth of 1kHz, this results in a sub-resonant, single frequency noise amplitude $N_{SF} \geq S_{SF}\sqrt{BW} \approx 3 - 6\ pm$. Depending on experimental details such as the reflectivity of the lever, roughness of the reflecting surface of the lever, beam size, position and shape, this number can worsen. If we assume $N_{SF} = 3pm$ noise floor and we limit our AC bias amplitude to $V_{therm}$, the piezoresponse of the sample material needs to be $d_{V_{ther}} = N_{oSF}/V_{therm} \approx 115\ pm/V$ for a signal-to-noise (SNR) of unity. Typically, one would like the SNR to be much larger than unity; that then requires either a much higher $d_{eff}$ or larger $V_{ac}$ bias. For OBD measurements, the spot position and size, as well as the resonant mode (frequency) play a large role in the sensitivity of the measurement. [18], [19] Essentially, the mode shape of the oscillating cantilever depends on the frequency, and since OBD measurements infer tip motion from a measured angle, the sensitivity values that depend on frequency in a non-trivial manner. Since the oscillation shape of a cantilever changes dramatically as the frequency ranges from near DC through the first contact resonance, there are enormous uncertainties in the calibration of levers as the frequency changes, an effect exacerbated by operating close to resonance. [20, 21], [22] Finally, note that OBD measurements are also vulnerable to crosstalk, where in addition to the vertical signal, long-range electrostatics, [21] ,[23] in-plane forces[24], [25] and inertial dynamics[26], [27] contribute to the measured amplitude and phase, complicating the interpretation. These results for single frequency, OBD detected (abbreviated "oSF") measurements are summarized in the first row to Table I below. We label the other techniques discussed in this paper as "oDART" for conventional OBD-detected DART,[28] "iSF" for interferometrically detected, single frequency[23] and "iDART" for the new interferometrically detected frequency tracking technique described in this work.

| Mode and Detector | Crosstalk | | | Accuracy | Noise amplitude ($N_k$) † | | $d_{k,V_{therm}}$ (pm/V) | | Features and Challenges |
|---|---|---|---|---|---|---|---|---|---|
| | $ES^1$ | $Dyn^2$ | $xz^3$ | | Ideal | Demonstrated | Ideal | Demonstrated | |
| Single frequency OBD ($k = oSF$) | ☑ | ☑ | ☑ | Variable | $3 - 6$ | 10 | 115 | 390 | Simple interpretation, component mixing, electrostatics and limited sensitivity |
| Resonance Tracking OBD ($k = oDART$)* | ☑ | ☐ | ☑ | Poor | $0.06 - 0.2$ | 0.5 | 6.2 | 19 | Has component mixing and electrostatics. Amplitude calibration very difficult. |
| Single frequency interferometry ($k = iSF$) | ☐ | ☐ | ☐ | Excellent | 0.16 | 0.5 | 5.8 | 19 | Spot above the tip |
| Resonance Tracking interferometry ($k = iDART$) | ☑ | ☐ | ☑ | Variable | 0.004 | 0.008 | 0.36 | 0.31 | Resonance gain + Very low noise detection. Calibration challenges mitigated by QPDI |

†Sampling frequency $= 1kHz$
[1]Body Electrostatic Forces
[2]Cantilever dynamics (resonance)
[3]Mixing of in-plane (x) and vertical (z) forces

*Table I Expected and measured noise performance for various detection and operating schemes discussed in the main text.*

An analysis for resonance-enhanced measurements needs to include a gain factor that is related to the resonance quality factor. The quality factors for typical ambient PFM cantilevers in this work are on the order of $Q_{contact} \approx 30 - 80$. Since oDART uses two frequencies on either side of the resonance peak, the actual gain is smaller than $Q_{contact}$ $G_{oDART} \approx 42$ for a $\Delta f = 6kHz$ (See Figure S4). This implies an oDART amplitude noise floor of $N_{oDART} = S_{oSF}\sqrt{BW}/G_{oDART} \approx 0.16pm$. Band Excitation using OBD (oBE) [29] is a closely related technique. The theoretical and experimental exploration of the noise floor for oBE is beyond the scope of this work, except to say that we have found the two to be comparable (see Fig. S2a). Converting this to the minimum converse piezo-sensitivity allows measurements at a thermal voltage, $d_{oDART,V_{therm}} = N_{oDART}/V_{therm} \approx 6.2pm/V$ - a substantial improvement over the single frequency approach.

While resonance-enhanced measurements greatly expanded the range of piezo- and ferro-electric material measurements, they are also subject to crosstalk effects, including electrostatics and mixing of in-plane and vertical contributions into the overall measured signal.[22] For electrostatics, since the tip–sample system is inherently capacitive, when a bias is applied, long-range electrostatic forces arise across the entire extent of the cantilever. These forces are not localized to the apex of the tip, but act as body forces along the entire length and can lead to displacements that can mimic piezoresponse at the tip-sample junction. [30] Especially at higher drive AC voltages or with weaker materials, these electrostatic contributions can dominate the signal, introducing contrast that is indistinguishable from electromechanical response, thereby greatly complicating the interpretation of PFM measurements.[31] Regarding component mixing, the electric field under the AFM tip excites both out-of-plane and in-plane displacements, and in-plane forces acting on the tip couple into the vertical PFM channel, especially in OBD measurements, where the detector measures cantilever angle rather than tip displacement. Furthermore, the degree of mixing is highly sensitive to the laser spot position along the cantilever. [24]

Clamping from stiff substrates, surface steps, or grain boundaries can further distort the measured displacement. [32] Such effects alter the apparent effective piezoelectric coefficient, making it difficult to separate intrinsic piezoresponse from topographic or boundary-induced signals. Finally, crosstalk from cantilever resonances and transfer functions can introduce frequency-dependent artifacts unrelated to the sample.[21, 23] Compounding these effects, many PFM studies report amplitudes in arbitrary units without calibration, obscuring quantitative comparisons.[31]

All three of these crosstalk effects – resonance artifacts, long-range electrostatics, and vector response – helped inspire the development and application of interferometric detection. [23, 24, 24] The single frequency approach to vertical PFM measurements with interferometry involves placing the detection spot over the tip and applying a single frequency, sub-resonant AC bias. We denote this approach as iSF. This approach benefits from the recently developed QPDI sensor, which has an amplitude noise density of $S_{iSF} \geq 5\, fm/\sqrt{Hz}$, implying an amplitude noise floor of $N_{iSF} = S_{iSF}\sqrt{BW} \approx 0.16pm$, similar to the oDART noise floor. Figure S2 shows a comparison of the contrast from oSF, oDART, iSF, and iDART at a range of drive amplitudes. Note that quantitative comparison between iSF, which is dominated by the vertical piezoresponse, and the other three approaches – oSF, oDART, and iDART - is complicated since they are sensitive to the mix of the vertical and lateral piezoresponse signals and electrostatic contributions.

In summary, oSF provides a baseline measurement approach but suffers from poor SNR and crosstalk; oDART offers high sensitivity at the cost of resonance-related artifacts; and iSF achieves comparable sensitivity to oDART while maintaining clean, artifact-free detection, making it the most direct and quantitative approach among the three. All of oSF, oDART, and iSF have amplitude noise floors that make reliable and high-fidelity measurements of weak piezoelectrics ($d_{eff} \leq\ 10pm/V$) problematic without requiring large and, in some cases, potentially destructive AC biases, $V_{ac}$. To mitigate this situation, we have combined low-noise, quantitative interferometric detection with the cantilever contact resonance (iDART). Using this approach, we demonstrate substantial SNR improvement, allowing reliable and non-destructive testing of weak piezoelectric materials at small AC biases.

# Part II. iDART Implementation and Demonstration

iDART leverages the femtometer-scale displacement sensitivity of quadrature phase differential interferometry (QPDI) with the resonance amplification provided by DART. Unlike conventional optical lever detection, QPDI directly senses cantilever displacement via phase shifts in reflected laser light. With an amplitude noise density of $S_{QPDI} \leq 5\ fm/\sqrt{Hz}$, it enables the detection of sub-picometer motion with an accuracy traceable to the wavelength of the light source.

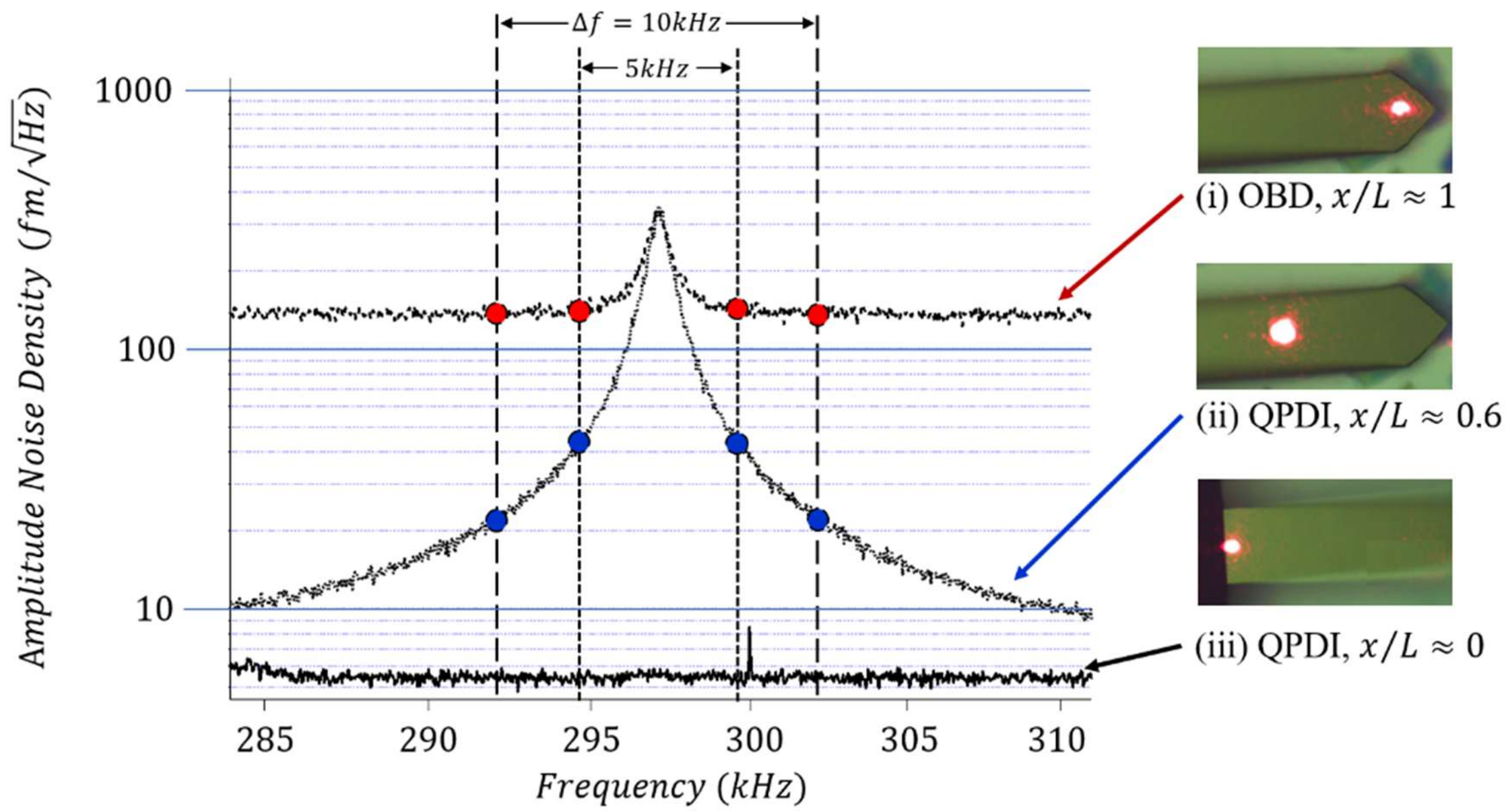

*Figure 1. Brownian noise spectra of a cantilever in contact with a surface, measured with OBD and QPDI. Details are discussed in the main text.*

Figure 1 shows three amplitude noise density spectra measured around the contact resonance of an Adama cantilever pressed against a surface: (i) the OBD measurements made at $x/L \approx 1$. The relatively high noise level of $\approx 150\ fm/\sqrt{Hz}$ only allows the very peak of the lever thermal resonance to be visualized. (ii) the measurements using QPDI at $x/L \approx 0.6$, showing the same thermal resonance and the spot position near the displacement maxima about halfway between the tip and the base. In this case, the noise floor is well below the Brownian (thermal) motion, and the entire resonance curve is visible. Finally (iii) the measurements using QPDI at $x/L \approx 0$ showing the fundamental noise limit of $\approx 5\ fm/\sqrt{Hz}$ of the QPDI detector.

In Fig. 1, it is notable that the detector noise floor is lower than the *off-resonance* thermal motion of the cantilevers we used in this work, a situation typically reserved for very small cantilevers in a fluid environment. This means that, while OBD measurements can resolve the thermal resonance peak above the noise floor, it does so only in a very narrow bandwidth. The red markers in Fig. 1 show OBD measurements that are separated by different values of $\Delta f$ that are limited by the noise floor of the OBD detection. Interferometric measurements, on the other hand, resolve the shoulders of the thermal peak at effectively an arbitrary bandwidth (see the blue markers in Fig. 1). For example, in DART frequency tracking, the two drive frequencies $f_{D1}$ and $f_{D2}$ should ideally be separated by $\Delta f \geq 2BW$, where $\Delta f = f_{D2} - f_{D1}$ and $BW$ is the measurement bandwidth. For many practical imaging situations, especially for faster scanning, this means $\Delta f \geq 5\ kHz$ as indicated in Fig. 1. In many cases, this means that the cantilever motion will be below the noise floor of the detector unless a relatively large excitation voltage is applied.

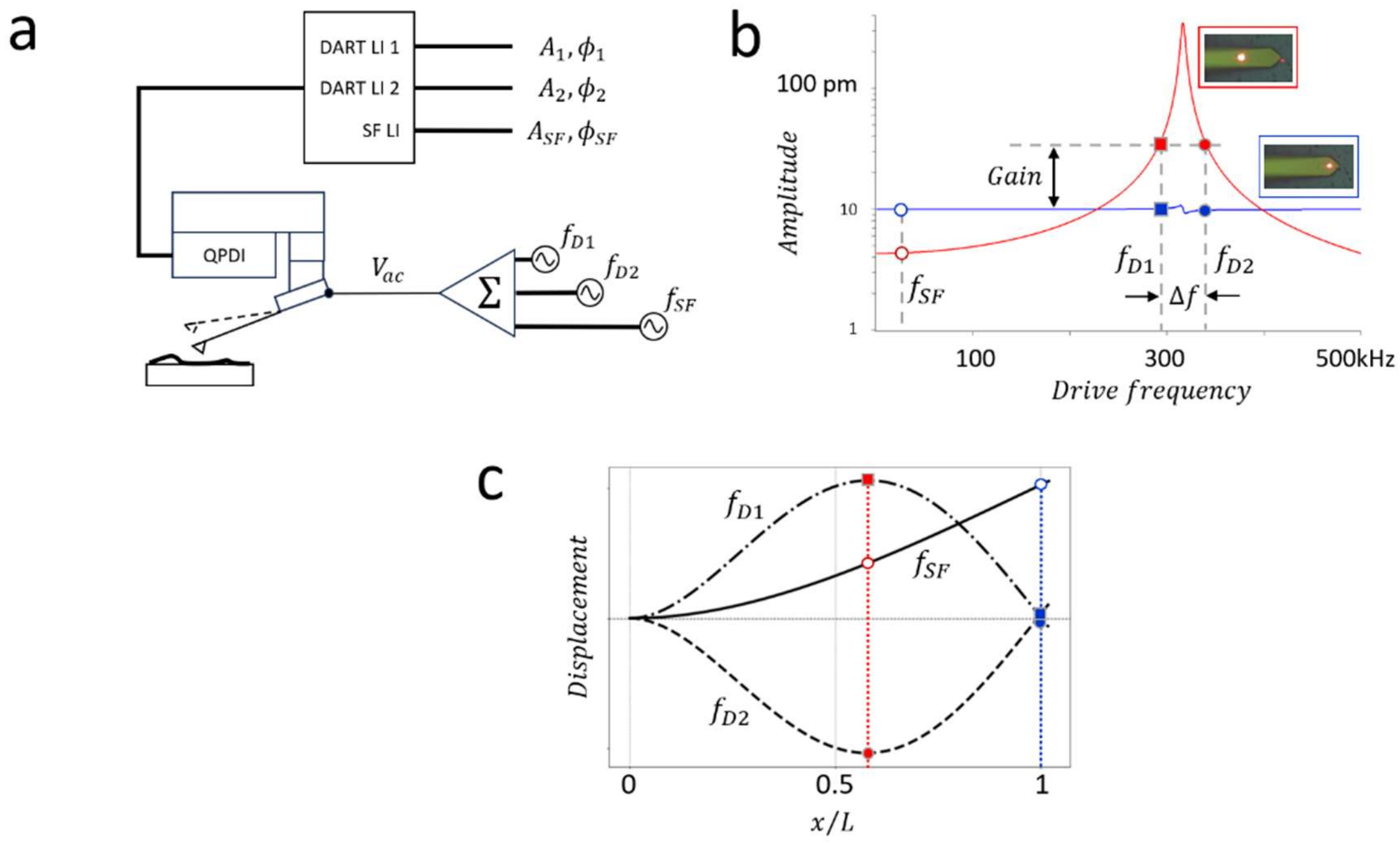

*Figure 2 (a) The experimental setup allowing simultaneous iSF (using $f_{SF}$) measurements and iDART (using $f_{D1}$ and $f_{D2}$). This arrangement generates simultaneous, pixel-to-pixel measurements, allowing easy comparison of iDART measurements with more conventional iSF measurements. (b) Schematic frequency response illustrating the frequency-dependent amplitude at two spot positions, over the tip at $x/L = 1$ (red) and over the body at $x/L$=0.6 (blue) (c) Cantilever displacement mode shape versus $x/L$ at the three frequencies; one at a sub-resonant $f_{SF}$ and two near the resonance frequency, $f_{D1}$ and $f_{D2}$. The amplitudes of the near resonance mode shapes were divided by 10x to allow easy comparison with the sub-resonant curve ($f_{SF}$). The sign change between $f_{D1}$ and $f_{D2}$ originates with the $180°$ phase shift of the lever as it passes through resonance.*

Figure 2a shows the iDART experimental setup used in this work. To both implement iDART and to simultaneously compare it to the conventional state of the art PFM, the QPDI signal is analyzed at three different frequencies. Two of these, $f_{D1}$ and $f_{D1}$, are applied on either side of the cantilever's contact resonance frequency and are separated by $\Delta f = f_{D2} - f_{D1}$, and used as the inputs into a DART feedback loop. The third frequency, $f_{SF}$, probes the response at a single, low sub-resonant frequency, allowing simultaneous, pixel-by-pixel comparison of iDART imaging with highly sensitive iSF measurements. Fig. 2b shows the amplitude frequency response with the spot at $x/L = 0.6$ (red) and at the tip $x/L = 1$ (blue). Note that at $x/L = 1$, the response remains essentially flat across frequencies.[23, 24] The inset optical micrographs of the lever with the spot location marked. Figure 2c shows a profile of the cantilever shape at the three frequencies as a function of position. When the detection spot is placed above the tip (blue curve), the response is nearly frequency independent. When the spot is located closer to the cantilever base (red curve), the low-frequency response is reduced because displacement depends on the relative spot position along the lever, while the resonance response is strongly amplified. We also anticipate that this general approach of combining interferometry with both force modulation[33] and contact resonance will benefit techniques other than DART, notably band excitation,[29] SPRITE[34] and related broad-band techniques.[35] We also anticipate that this will improve other application areas, notably nanomechanical measurements that make use of mechanical[36], [37], [38] or photothermal actuation [39] that may benefit from both small amplitudes and interferometric calibration.

As a first example, we applied iDART to ferroelectric films of Y:$HfO_2$ (see Materials and Methods for details). These films are known for their weak piezoelectric properties ($d_{eff}$ of about several pm/V). These films were chosen because, in spite of their ferroelectric nature confirmed by polarization hysteresis measurements, they typically do not show clear evidence of ferroelectric domains using conventional oSF, oDART, or iSF approaches discussed in Part I, and as shown in Figure 3.

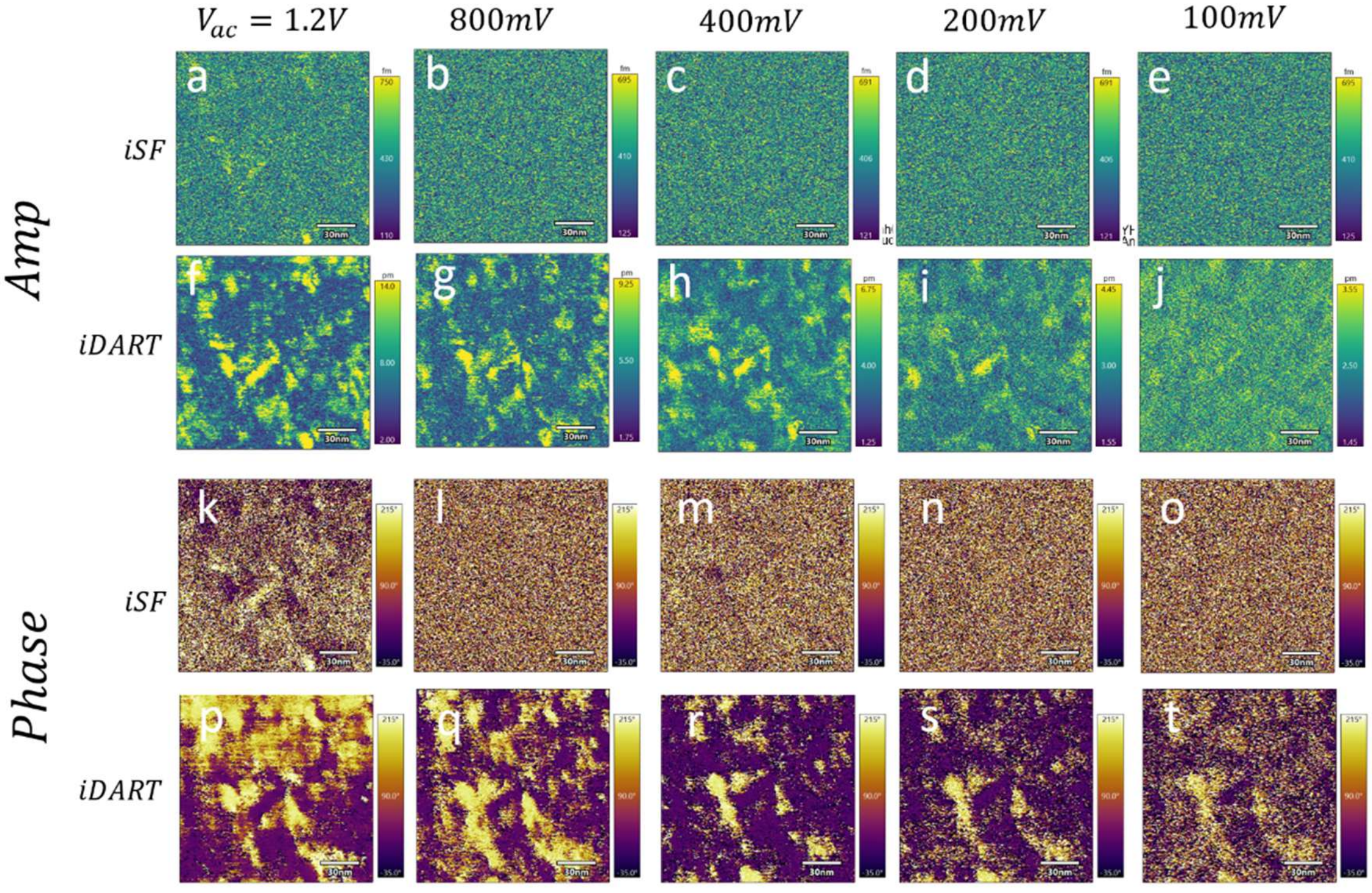

*Figure 3. Simultaneously acquired iDART and iSF PFM images of the Y:HfO$_2$ films. Panels a-e show the iSF amplitude images as the AC bias ranges from 1.2V (a) to 100mV (e). Panels k-o show the associated iSF phase images. Panels f-j show the iDART amplitude images, and panels p-t show the associated iDART phase images. Histograms for the iSF and iDART amplitude and phase images are shown in supplemental Fig. S8. The average $d_{eff}$ for the measured with this probe and calibrated with iSF was $d_{eff}$~0.2 pm/V, conservatively 100x times weaker than in many conventional perovskite ferroelectrics (see data for PZT films in Fig. 5).*

In the images of Y:HfO$_2$ films shown in Fig. 3, the iSF amplitude (a–e) shows weak contrast at the largest AC bias (a) and noisy contrast across all drive levels, with domain structure essentially unresolved even at the highest bias of 1200 mV and below the noise floor at lower AC bias values. Furthermore, there was a clear modification of the sample surface when imaging at higher biases, confounding interpretation and leading to irreversible modification of the sample polarization and physical structure. On the other hand, the iDART amplitude images (f–j) consistently showed contrast that persists down to $V_{ac} = 100mV$. Similarly, while the iSF phase (k-o) had little discernible signal for $V_{ac} < 1200mV$ (see panels b-e) and the iDART phase images maintain clear contrast down to $V_{ac} = 100mV$. Notably, the two higher bias images show an increase in the regions with the larger phase values (see the growing number of yellow pixels in panels q ($V_{ac} = 800mV$) and p ($V_{ac} = 1200mV$). This is consistent with the sample polarization evolving under the influence of the large AC bias, whereas the images acquired using small biases remain much more stable. This also implies that the only iSF image with contrast (panel a, $V_{ac} = 1200mV$) comes at the cost of irreversible modification of the very structure being imaged.

Amplitude and phase histograms for the images in Fig. 3 are shown in the Supplemental Figs S8 and S9. In particular, the iSF and iDART phase histograms obtained from images in Fig. 3(k-o) and (p-t), respectively, indicate that the iSF phase distributions never clearly show the expected 180-degree bimodal distribution, while the iDART histograms maintain narrow, bimodal separation of 180 degrees, as expected from the piezoresponse of a ferroelectric material.

Overall, these results demonstrate that iDART outperforms iSF and oDART by maintaining robust amplitude and phase sensitivity in Y-doped HfO$_2$ thin films, enabling reliable imaging of weak piezoelectric signals at excitation levels an order of magnitude lower than before.

iDART can be used to perform switching spectroscopy PFM (SSPFM) as well, where the dc bias is modulated at a low frequency (typically ramped or stepped). In this work, we used the stepped ramp developed by Jesse et al. [31,40] where the stepped DC bias is applied while the “on” response is measured and then removed, allowing the “off” (remnant) response to be measured.

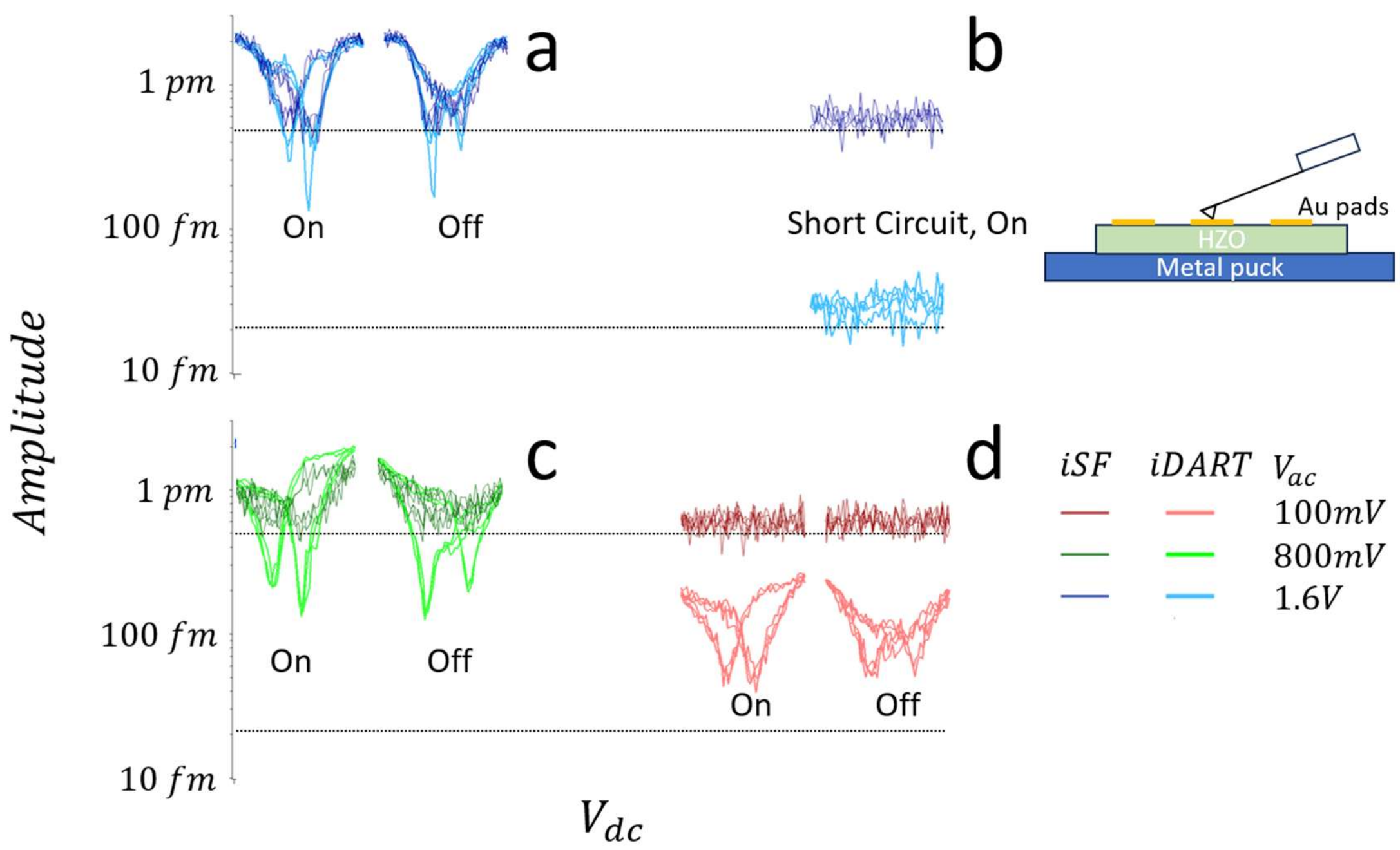

*Figure 4. Switching spectroscopy "on" and "off" amplitude butterfly loops measured on the 5-um-diameter 5-nm HZO capacitors at different AC biases; 1.6 V (blue traces), 800 mV (green traces) and 100 mV (red traces). The dark blue, green and red are the iSF data, simultaneously acquired with the light blue, green and red iDART data. The amplitudes presented have been corrected for resonance gain, described in Gannepalli et al.* [22] *and discussed in the text.*

The switching spectroscopy amplitude measurements shown in Fig. 4 compare iSF and iDART responses under varying AC drive conditions. At high drive (1.6 V, blue traces), both iSF (dark blue) and iDART (light blue) probing produced clearly resolved butterfly loops with large amplitudes. The device failed during the 1.6V high bias acquisition (see supplemental Fig. S1 for the time sequence data). This allowed us to acquire baseline noise measurements on the short-circuited device. The iSF short-circuit amplitude was ~0.5 pm while the iDART short-circuit amplitude was ~25 fm, representing the noise floor of the two measurement approaches.

By lowering the AC biases (800 mV and 100 mV), we were able to avoid electrical breakdown. However, the iSF amplitude loops were only barely visible at 800mV drive amplitude (dark green) and disappeared below the noise floor (dark red) at 100mV. In contrast to the iSF results, iDART curves clearly resolved both the on and off butterfly loops at both the 800 mV and 100 mV drive bias values (light green and red, respectively). By leveraging resonance enhancement and lower noise, iDART reduces the required magnitude of $V_{ac}$ by more than a factor of 10, allowing stable and reproducible ferroelectric switching measurements.

The results of a study comparing simultaneous iDART and iSF measurements on a polished $PbZrTiO_3$ (PZT) sample are shown in Fig. 5. We chose this sample because of its large expected $d_{eff}$, allowing conventional measurement modes to map the domain structure easily. At the same time, by decreasing the magnitude of $V_{ac}$, we can explore relative noise thresholds with different imaging modes.

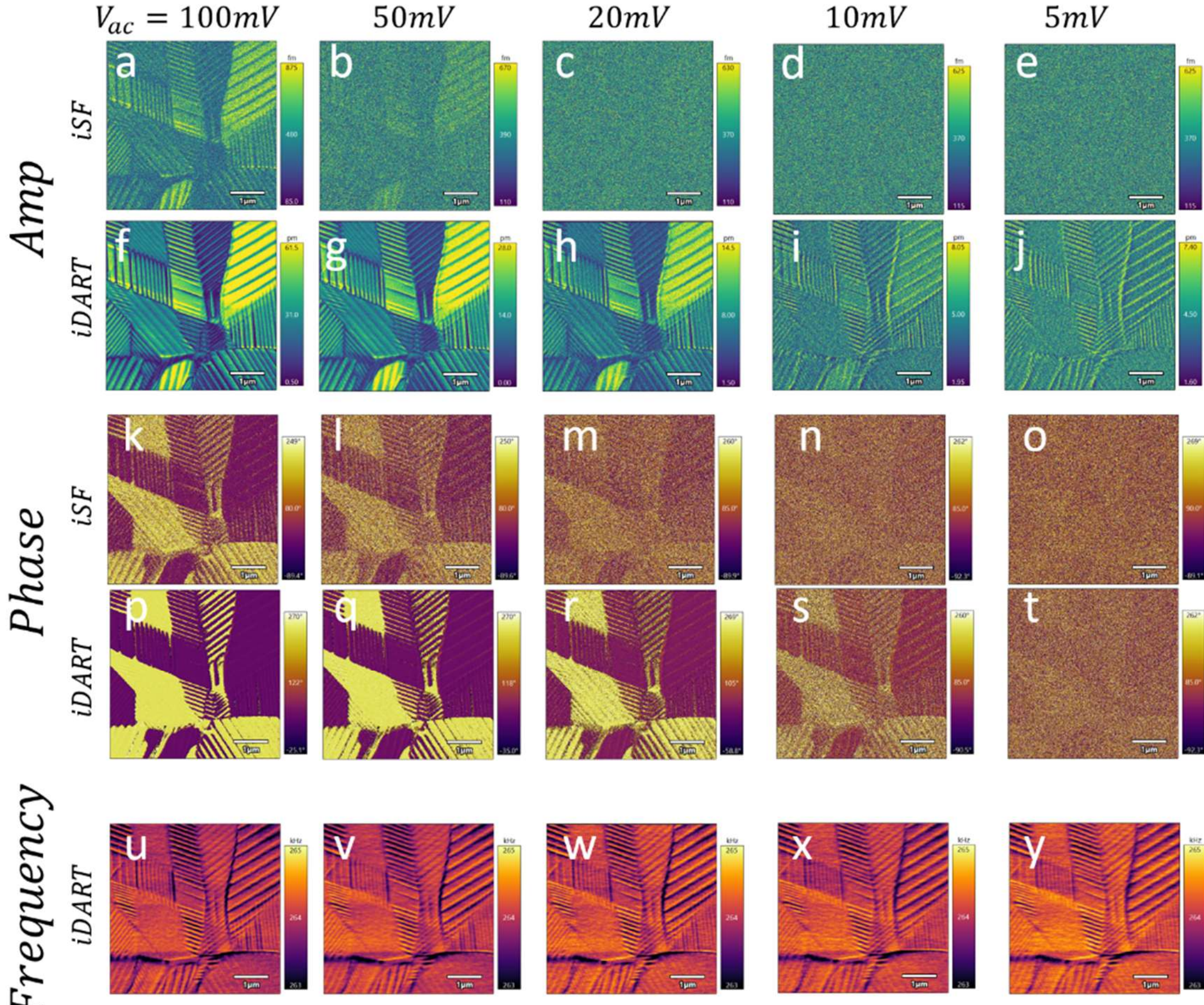

*Figure 5. Simultaneously acquired iSF and iDART PFM images of PZT. Panels (a-e) show the iSF amplitude images, acquired at a drive frequency of 42 kHz and (f-j) show the simultaneously acquired iDART amplitude images for comparison. The AC bias ranged from 100 mV to 5 mV. Panels (k-o) show the iSF phase images, and (p-t) show the iDART phase images. u-y Panels (u-y) show the iDART contact resonance frequency images. All of the iSF images were acquired at a drive frequency of 42 kHz. The associated histograms of the amplitude, phase, and contact resonance images are shown in supplemental Fig. S9.*

In Fig. 5a-e, the iSF amplitude images (a–e), the domain contrast is strong at 100 mV (a) but rapidly degrades as the drive is reduced. At 50 mV (b), the signal is already weak, and by 20–5 mV (c–e) the amplitude disappears below the noise floor, leaving domains invisible. In contrast, the iDART amplitude images (f–j) preserve clear, sharp domain contrast even at low drive levels. Although the absolute signal decreases somewhat with reduced bias, the spatial contrast remains robust across the entire range down to 5 mV. This difference is quantified in the amplitude histograms in supplemental Fig. S9(a): the iSF distributions (dashed lines) shift into the noise floor as bias is reduced, while the iDART histograms (solid lines) remain well-separated, and distinctly above background at all drive levels.

The iSF phase images in Fig. 5k-o similarly show good contrast at higher bias (k and l) but lose clear domain definition at lower drive voltages (m-o), where the phase begins to blur into noise. By comparison, the iDART phase images (p-t) consistently display clear 180° contrast until the drive bias drops to 10 and then 5 mV $V_{therm}$ (s-t), presumably, where the piezoresponse amplitude starts to be comparable to the Brownian motion of the cantilever. Supplemental Fig. S9 shows the histograms associated with Fig. 5 and reinforces this point: iSF distributions broaden and flatten at low bias, while iDART maintains two narrow, well-separated peaks separated by $\sim$180° that merely broaden as the drive amplitude is reduced and therefore the signal-to-noise decreases. The iDART measurements have the added benefit of providing nanomechanical information about the tip-sample contact stiffness through the contact resonance frequency. For example, in this case, the tip–sample contact was highly stable with a nearly constant CR frequency shown in the images (u-y). The associated histograms presented in supplemental Fig. S9(c) show that there is a slight systematic increase in contact stiffness, possibly indicating a modification in the sample surface or tip wear. This highlights the usefulness of obtaining simultaneous PFM amplitude and phase information along with localized nanomechanical stiffness information.

In SS-PFM, the coercive bias extracted from a local hysteresis loop is an operational quantity rather than a purely intrinsic material constant, because the measured loop depends on the tip-induced nucleation barrier, domain-growth kinetics, and the details of the applied waveform and probing conditions. Consequently, comparisons of coercive bias are only meaningful when bias amplitude, pulse timing, and on-/off-field measurement conditions are controlled, since finite probing bias and electrostatic offsets can themselves modify loop width and horizontal position (1.1, 2.1, 3.1). In this context, increasing the applied bias amplitude can make the apparent coercive bias smaller because the stronger local electric field reduces the activation barrier for domain nucleation and accelerates subsequent domain-growth kinetics, so switching is triggered earlier during the voltage sweep rather than at the small-signal threshold (4.1, 1.1, 2.1). At the same time, large-signal excitation can alter the loop shape through non-ferroelectric contributions: electrostatic tip-sample forces can introduce horizontal offsets related to surface potential and finite probing bias, while charge injection, surface charging, current flow, and associated Joule heating can broaden, skew, or abruptly shift the loop and therefore bias the extracted coercive voltage to lower values (5.1, 5.2, 5.3, 3.1). More broadly, PFM hysteresis can also be affected by carrier/ion migration, electrochemical processes in the tip-surface junction or water meniscus, and high-order nonlinear strain mechanisms, all of which may generate imprint-like shifts or ferroelectric-like hysteresis that complicates the interpretation of the measured coercive bias under high-bias conditions (6.1).

Larger bias amplitudes can artificially reduce the apparent coercive bias because the stronger electric field lowers kinetic barriers and accelerates domain/nucleation dynamics, so switching occurs earlier in the sweep. In addition, large-signal excitation can introduce field-driven nonlinearity and heating/charge-injection effects that broaden or distort the response and shift the loop horizontally, making the extracted coercive bias appear smaller than its small-signal value. Figure 6 summarizes the piezoresponse (PR) loop area, normalized by the ac drive amplitude ($\int \mathrm{PR}\, dV_{\mathrm{bias}}/V_{\mathrm{ac}}$, where $PR = A \cdot \cos\phi$) for two different samples: (i) fused silica and (ii) 9-nm-thick Annealed $HfO_2$ capacitors. Supplemental Fig. S6 shows the loops that were used to derive the areas in Fig. 6. Referring to Fig. S6, the piezoresponse (PR) loops of the annealed $HfO_2$ films exhibit a coercive voltage $V_{\mathrm{coercive}} \approx 3$ V. This correlates well with the roll-off in the normalized loop area observed in both the PR-on and PR-off panels of Fig. 6. At this highest drive amplitude, the $HfO_2$ loop area converges to values comparable to that of fused silica, indicating a suppression of ferroelectric-like hysteresis at large excitation amplitudes. Notably, however, the reduction in loop area for $HfO_2$ begins at substantially lower drive amplitudes, between approximately at $V_{\mathrm{ac}} = 400 - 800\ mV$. This is on the order of $16 - 32$ thermal voltages ($V_{therm} = k_B T/e$). Additionally, the $HfO_2$ loop area decreases with increasing applied load. While the underlying mechanism requires further investigation, this trend may reflect effects including (i) increasing the contact area with higher loads means that the tip samples a larger population of disordered polar nanodomains, thereby reducing the over-all signal or (ii) pressure-induced clamping that suppresses the piezoresponse. In contrast, fused silica exhibits intrinsically small piezoresponse signals and correspondingly low hysteresis loop areas across all loads. For these materials, the normalized integrated loop area remains approximately constant over the full range of $V_{\mathrm{ac}}$, consistent with a largely non-hysteretic, electrostatic or linear electromechanical response rather than with ferroelectric switching.

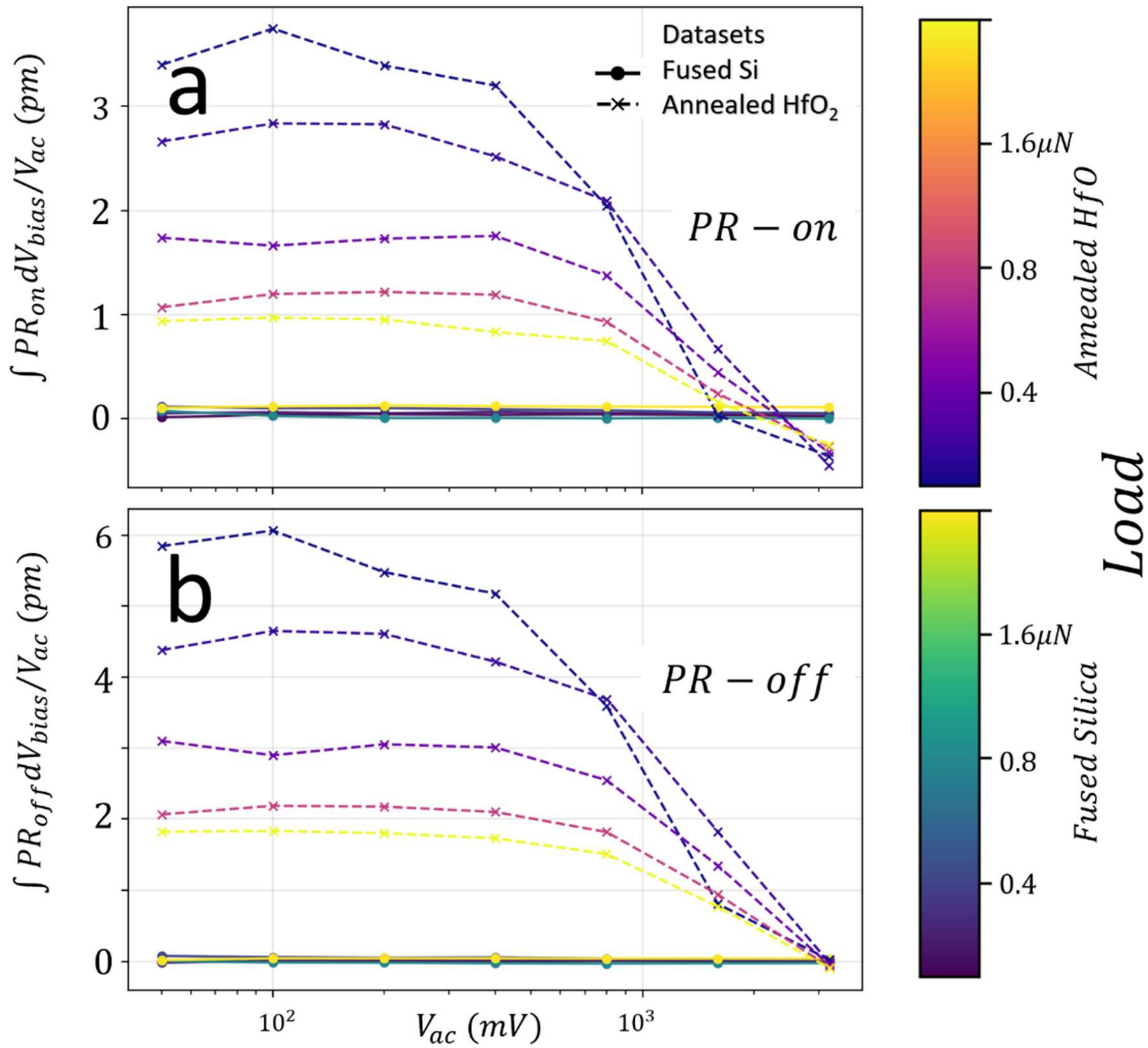

*Figure 6. Normalized hysteretic piezoresponse loop area for a 6 nm thick, annealed HfO capacitor structure (crosses, dashed lines) as a function of both AC drive amplitude and loading on the sample. Loop areas for the reference fused silica (squares, dash-dot lines) are also included. (a) the PR-on, and (b) PR-off normalized loop areas plotted versus AC drive amplitude $V_{ac}$ for the two samples. Curves are color-coded by applied normal load (see color bars at right), using the same series of loads for all the materials and the same cantilever. The normalization by $V_{ac}$ removes trivial scaling with excitation amplitude, enabling comparison of the intrinsic hysteretic electromechanical response across samples, loads, and drive conditions. See supplemental Fig. S6 for the PR loops used to generate these data.*

For over 20 years since its development, a major challenge with oDART PFM has been quantifying the amplitude response. There are two components to this: the amplifier gain, related to the Q-factor, and the OBD sensitivity calibration. An early attempt was made to quantify oDART response in the context of a simple harmonic oscillator (SHO) model, accounting for the quality factor of the oscillator.[22] However, this approach does not take into account unknown changes in the OBD sensitivity as the cantilever mode shape evolves from sub- to near-resonance. [18,] In contrast, since the interferometer sensitivity is determined by the wavelength of light, calibration of iDART sensitivity may be more easily interpreted. An example is shown in supplemental Figs S5d-f, where we compare iDART $A_{iDART}$ measurements made with $V_{ac} = 100\ mV$ (unmasked, S5a and masked, S5d) with iSF $A_{iSF}$ measurements made at $V_{ac} = 1\ V$ (unmasked, S5b and masked, S5e). The ratio of these two measurements, $G = (A_{iDART}/V_{ac,iDART})/(A_{iSF}/V_{ac,iSF})$ is plotted in Fig. S5c, providing an estimate of the resonance amplification $G \approx 33.8$ (see the red and green curves in Fig. S5g). We expect substantial variations in gain since it is well established that the resonance quality factor varies from pixel to pixel.[22, 39]

We can use this gain ratio to estimate the noise floor of iDART measurements. For the cantilever used here, the measured noise amplitude at the spot location $x/L \approx 0.6$ is $A_{iDART,\, x/L=0.6} \approx 3\ pm$. This implies an input noise level of $N_{iDART,\, x/L=0.6}/G \approx 10\ fm$, corresponding to a noise density of $S_{iSF,\, x/L=0.6}/G \geq 0.3\ fm/\sqrt{Hz}$. This in turn, implies that materials with an effective converse piezo sensitivity of $d_{V_{iDART,\ ther}} = N_{iDART,\, x/L=0.6}/V_{therm} = 0.36\ pm/V$ should have an SNR~1 when excited at the thermal voltage, $V_{therm} = 25.7\ mV$. These results are summarized in the last row of Table 1 for iDART and the other measurement modes. We anticipate that there is additional room for improvement in the iDART measurements, for example, by further optimization of the spot location or using smaller cantilevers with reduced damping and therefore reduced thermal noise, as has been exploited in other measurements.[41, 42]

Finally, we note that while iDART dramatically improves the sensitivity of electromechanical measurements, since the spot location $x/L \neq 1$, it is subject to similar crosstalk challenges that have been discussed at great length elsewhere. Explicitly, as with oSF, oDART and oBE, both electrostatics and in-plane (longitudinal) forces will mix with the vertical response, so caution should be exercised in both experimental design and interpretation of the results. However, the drastically enhanced sensitivity of iDART allows operation at substantially lower modulation voltages, in some cases well below the thermal voltage. Operating at such low modulation voltages suppresses, rather than amplifies, many parasitic effects that are otherwise exacerbated at high modulation voltages, including electrostatic crosstalk, Joule heating, ion migration, and charge injection. The latter two are the most serious mechanisms affecting the genuine, i.e., ferroelectric-related, PFM signals and may well be reduced to negligible levels, meaning that the high sensitivity of iDART will actively suppress those parasitic effects.

The accuracy of PFM measurements are limited by (1) short-range electrostatic coupling between the tip and the sample (resulting in a localized, electrostatic modulation force), (2) long-range electrostatic coupling between the cantilever body and the sample, [21, 5, 24, 30] and (3) cantilever transfer function variations, especially for single-frequency PFM in the presence of variations in the tip-sample stiffness (and, therefore, the cantilever resonance frequency and quality factor).[26], [27] Interpreting the measured signal without considering these factors can lead to serious interpretational errors and is likely a common reason for erroneous reporting of nanoscale piezo- and ferroelectricity. [31]

Operating at low drive frequency can reduce electrostatic amplification and improve quantitative reliability. However, "low frequency" is not a universal condition; it depends strongly on the magnitude of the electrostatic contribution and may, in some cases, need to be well below a few kHz. Conversely, resonance-enhanced PFM offers an improved signal-to-noise ratio but requires caution, because changes in local dissipation alter the quality factor and hence the gain of the resonance amplifier. Apparent variations in PFM amplitude may therefore reflect changes in resonance enhancement rather than true piezoelectric response.

Probe design provides another route for reducing artifacts. Smaller cantilevers and longer tips decrease the effective capacitance between the cantilever body and the sample, thereby reducing distributed electrostatic forces. Shielded probes may further suppress stray coupling, although they are currently more expensive and less developed than conventional cantilevers. Stiffer cantilevers can also reduce deflection from long-range forces, but their use may be problematic for thin films or soft materials because higher loading forces can damage the sample or alter its response.

Detection and measurement geometry should also be optimized. Placing the optical beam deflection spot closer to the cantilever base can reduce sensitivity to nodal structure in the cantilever motion, although this decreases detection sensitivity. Scanning near sample edges or measuring at different sample rotation angles can help identify nonlocal electrostatic contributions by changing the cantilever–sample coupling geometry.[21] Finally, and perhaps critically, the use of a top electrode can substantially improve the reliability of PFM measurements. An electrode stabilizes the sample surface, reduces sensitivity to humidity and surface contamination, and shields the tip–sample junction from localized non-piezoelectric electrostatic forces. In this way, electrode-based measurements provide an important control for distinguishing intrinsic electromechanical response from spurious electrostatic artifacts that can arise on an exposed surface.

Figure 7 illustrates this effect by comparing local hysteresis loops measured both on and off electrodes for an amorphous, non-ferroelectric, and an annealed, ferroelectric 6-nm-thick pure $HfO_2$ film. Macroscopic measurements on 6-nm-thick $HfO_2$ films show a weak ferroelectric switching peak resulting in $2P_r$ values of 4 μC/cm². Accordingly, a hysteretic PFM response should be expected. A sample with very low remanent polarization was selected to highlight the method's sensitivity.

In both materials, the off-electrode amplitudes are substantially larger than the on-electrode amplitudes, demonstrating that the exposed surface produces a much stronger apparent PFM response. This contrast is especially important for the amorphous sample, where no true ferroelectric switching is expected.

The ferroelectric annealed $HfO_2$ sample shows clear hysteresis, both on and off the electrode (Figure 7a). The bare annealed region exhibits the largest loop amplitudes, and both regions show clear hysteresis, consistent with a genuine electromechanical contribution from the ferroelectric sample. On the non-ferroelectric amorphous sample (Figure 7b), loops exhibit large hysteresis on the bare sample surface (top plot), inconsistent with ferroelectric tester results. Measurements acquired on the electrode are essentially non-hysteretic and show only a weak amplitude trend (bottom plot). The apparent electromechanical response measured on the electrode is approximately 5 fm/V, roughly 100 times smaller than that measured on the bare annealed ferroelectric region. This strong suppression indicates that much of the large bare-surface signal in the amorphous region is unlikely to originate from intrinsic piezoelectric or ferroelectric behavior. These results, particularly the loops of the bare amorphous region in Fig. 7b, emphasize that PFM loops on bare samples should be interpreted with high caution. The top electrode acts as a filter for spurious effects: it presumably suppresses surface charging, adsorbate-mediated forces, and localized electrostatic interactions while preserving measurable response in genuinely ferroelectric samples.

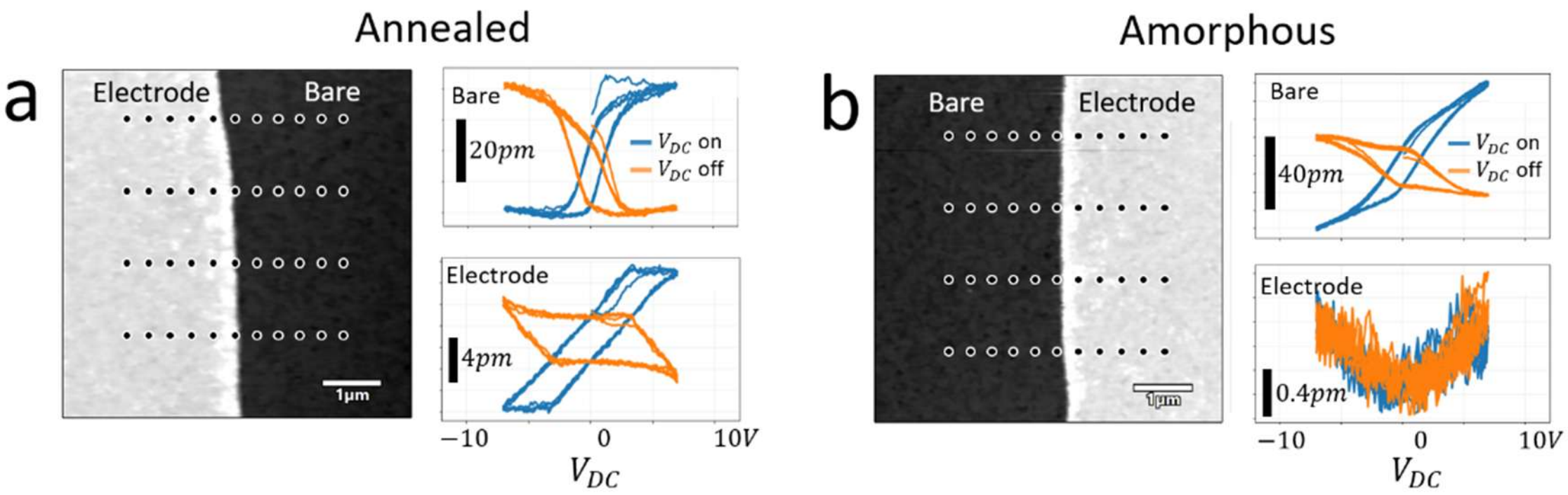

*Figure 7. Electrode-dependent iDART PFM response in annealed ferroelectric and amorphous non-ferroelectric $HfO_2$ films. (a) Topographic image of an annealed, ferroelectric $HfO_2$ sample, spanning the boundary between electrode-covered and bare areas, with markers indicating locations where local iDART PFM hysteresis loops were acquired. Loops averaged over the bare region of the sample (top) show a large switchable response on the bare surface and a smaller but measurable response in the loops averaged on top of the electrode (bottom). (b) Topographic image of a non-ferroelectric amorphous $HfO_2$ region spanning the boundary between bare and electrode-covered areas, with corresponding averaged hysteresis loops. In the amorphous film, the bare region exhibits a large apparent hysteretic response, whereas the electrode-covered region shows a near-zero displacement response. Blue and orange curves correspond to measurements acquired with the DC bias applied (VDC on) and removed (VDC off), respectively. The strong suppression of the amorphous-film response by the electrode indicates that the bare-surface signal is dominated by non-ferroelectric electrostatic or charge-mediated contributions, while the retained response in the annealed film is consistent with genuine ferroelectric electromechanical activity. Scale bars in the topographic images are 1 μm. Vertical scale bars in the hysteresis plots indicate the displacement amplitudes shown; the AC drive amplitude was 100 mV for all loops.*

The origin of the non-ferroelectric loops on the bare, amorphous $HfO_2$ surface is not currently clear, but similar phenomena have been observed in ambient environments in a variety of other systems, such as silicate glasses as well as other non-piezoelectric materials, for example in reference 5 and for the mica and soda-lime glass shown in Figure S10. On a pragmatic note and for the purposes of this discussion, since the overall ion-density is lower on the fused silica, we expect, and indeed observe, that the background on that substrate is systematically lower and so it offers a preferable electromechanical “null” reference material that we used in Figures 6 and S6.

Finally, it is important to note that contact-resonance (CR) measurements are useful well beyond electromechanical mapping. Tracking one or more cantilever resonances in contact enables quantitative, high-resolution maps of contact stiffness and elastic modulus across heterogeneous materials (metals/ceramics, composites, coatings, and multiphase polymers). With appropriate modeling of resonance frequency and dissipation (e.g., Q or damping), CR can also provide viscoelastic parameters such as storage and loss moduli, important for polymer blends, thin films, and other dissipative systems. ,[39], [43] Resonance frequency, amplitude, phase, and dissipation contrast can further reveal subsurface or embedded features through near-surface mechanical heterogeneity. [44] Contact resonance is also central to chemical mapping with photothermal Infra-red imaging (PTIR), where resonance enhancement transduces photothermal expansion into measurable motion.[45], [46] In this broader CR-AFM setting, iDART should improve simultaneous measurements of the contact resonance and dissipation by combining low-noise, calibrated interferometric displacement detection with robust DART, reducing the drive levels required to maintain bandwidth and sensitivity in lossy or weakly responding contacts.

# Conclusions

A fundamental limitation of conventional PFM stems from the need to apply large AC biases to overcome displacement detection noise. Large biases can induce electrostatic interactions that mimic piezoresponse, initiate surface chemical reactions, cause Joule heating, and tip-induced switching, all of which distort or, in some cases, destroy the localized electromechanical responses being studied. The problem is particularly severe for emerging ferroelectrics, such as hafnia-based thin films, antiferroelectrics, and 2D ferroelectrics, where intrinsic displacements are weak and bias-induced damage thresholds are small.

iDART overcomes these limitations by combining femtometer-resolution interferometric detection with resonance-enhanced dual-frequency tracking. The result is a >10x improvement in signal-to-noise over existing techniques, enabling reliable imaging and spectroscopy with mV-scale AC biases.

iDART measurements of the Y:$HfO_2$ films, known for their weak piezoelectric activity ($d_{eff} <$ 1 pm/V), confirm that iDART succeeds in revealing domain contrast where conventional methods fail. Specifically, pristine domain structure was visualized in the Y:$HfO_2$ films using AC biases as small as 100 mV. Switching spectroscopy further demonstrated that iDART resolves hysteresis loops in hafnia-based ferroelectric capacitors at much lower drive voltages (~100 mV) than the >1 V biases used in existing PFM approaches, alleviating the problem of device failure during testing.

In experiments with PZT ceramic samples, iDART maintained robust domain contrast down to 5–10 mV of AC bias, roughly an order of magnitude lower than required for conventional PFM with the same cantilever.

This high sensitivity establishes iDART as a powerful tool for reliable, non-destructive, quantitative, and reproducible electromechanical measurements of fragile and weak piezoelectric systems spanning ultrathin films, 2D materials, and beyond-Moore electronic devices.

# Acknowledgments

Work at the University of Nebraska was supported by the National Science Foundation (NSF), grant DMR-2419172. A.G. and X.X. acknowledge support of Intel Corporation. The authors would like to thank Dr. Haidong Lu and Mr. Amit Kumar Shah for their help with the preparation and characterization of the YHO films. F.W. was financially supported by the Deutsche Forschungsgemeinschaft (DFG) within the project D3PO (505873959). U.S. was supported out of the Saxonian State budget approved by the delegates of the Saxon State Parliament.

# Materials and methods

iDART was performed on a Vero interferometric (QPDI) AFM with custom code. Using software version 21.12.79, with some modifications enabling the QPDI signals to be used in the built-in DART feedback loops, as well as the Z servo loop.

The tip-sample amplitude can be estimated from the observed amplitude and phase observables using the method described in Gannepalli et al.[22] This method is integrated into the software. This approach was used for the hysteresis loops shown in Figures 4, 7.

PZT-5H polycrystalline ceramic sample several millimeters thick was first cut and mounted on epoxy, then sequentially polished with progressively finer silicon carbide abrasive paper under water. The final surface finishing was performed with diamond suspensions, 4,000 to 8,000 grit.

The epitaxial 5% Y-doped 7-nm-thick $HfO_2$ (111) thin films were grown on the $La_{0.7}Sr_{0.3}MnO_3/SrTi_{O3}$ (LSMO/STO) by pulse laser deposition (PLD) using a laser wavelength of 248 nm with a repetition rate of 2 Hz. The oxygen pressure was 70 mTorr and the substrate temperature was 730°C. At the end of deposition, the temperature of the films decreases to room temperature with a cooling rate of 10°C/min under an oxygen pressure of 70 mTorr.

The metal-ferroelectric stack was deposited on a p-doped silicon (p-Si) substrate. The bottom electrode (BE) contact was formed using direct current (DC) sputtering in a Bestec ultrahigh vacuum system with 30 nm of tungsten (W) and 10 nm of titanium nitride (TiN) at room temperature. Then, a 5-nm $Hf_{0.5}Zr_{0.5}O_2$ layer and a 6-nm pure $HfO_2$ was deposited on the BE via atomic layer deposition (ALD) using $CpHf[N(CH_3)_2]_3$ and $CpZr[N(CH_3)_2]_3$ as the Hf and Zr sources, respectively, in an Oxford OPAL ALD tool. For $Hf_{0.5}Zr_{0.5}O_2$ binary oxides, the metal precursor was alternated between Hf and Zr to achieve the desired $Hf_{0.5}Zr_{0.5}O_2$ composition. The top electrode (TE) was deposited in the Bestec ultrahigh vacuum system with a 10-nm-thick TiN film. Then, rapid thermal annealing was performed at 500 °C for 20 seconds. Finally, the TiN top electrode was etched using inductively coupled plasma (ICP) etching.

Supplemental Material

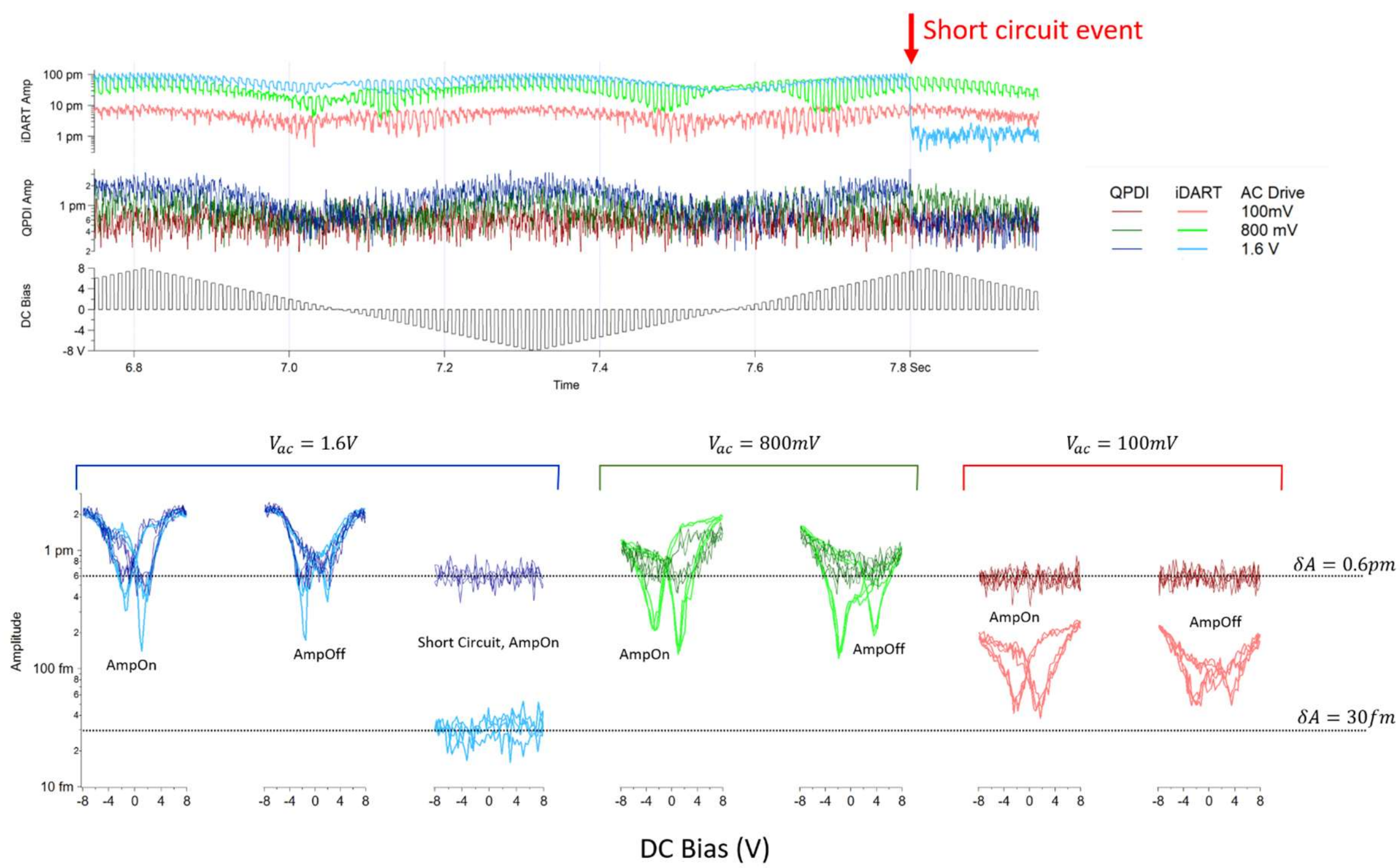

Figure S1.Time SSPFM data from Fig. 3. The short circuit even at large bias is visible at around the t=7.8second mark, near the largest positive DC bias value (~8 volts).

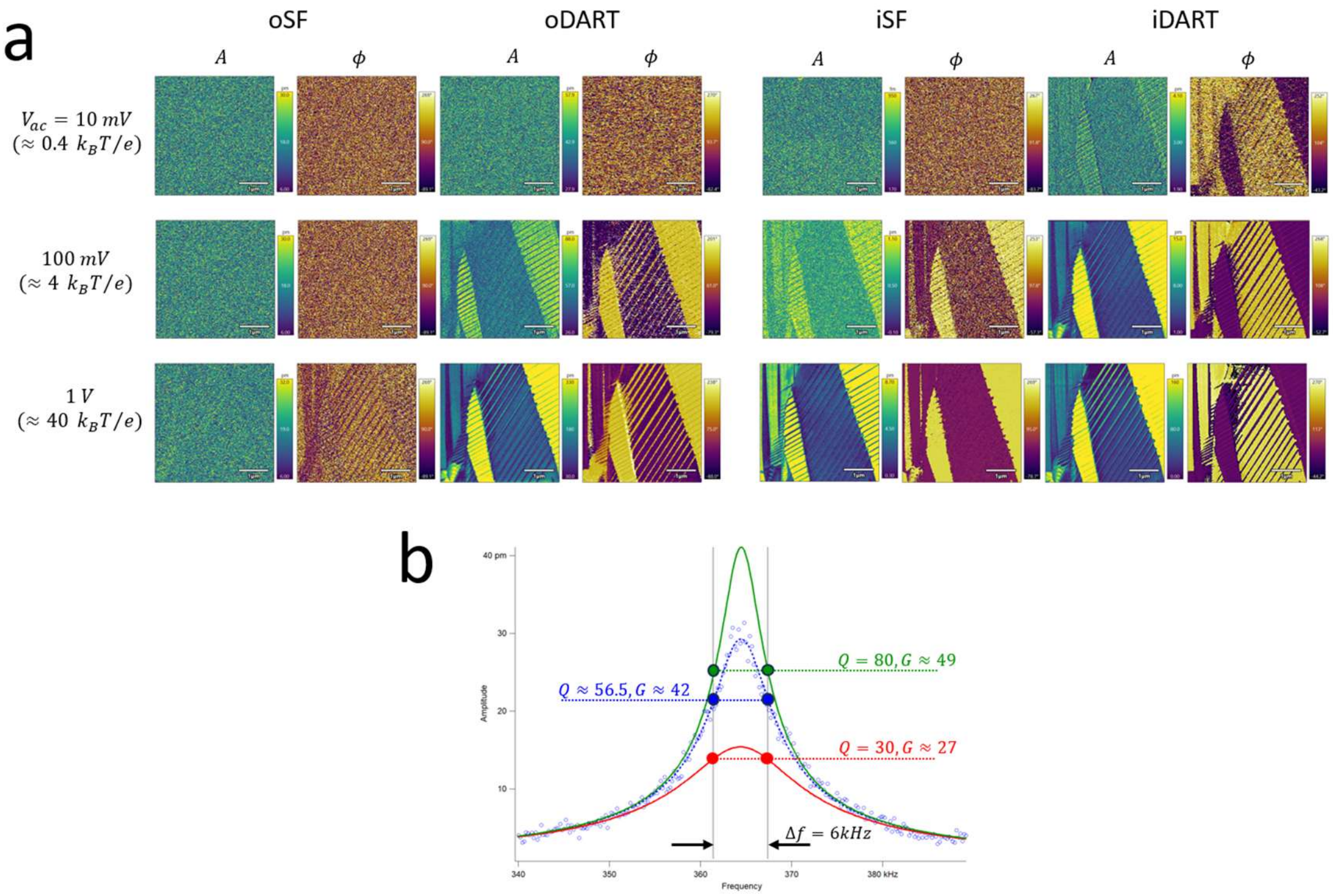

Figure S2. Comparison of piezoresponse imaging modes on PZT across drive amplitudes. (a)

Simultaneously acquired amplitude (A) and phase (ϕ) images of PZT using four measurement configurations—outer-lever simple frequency (oSF), outer-lever dual AC resonance tracking (oDART), inner-lever simple frequency (iSF), and inner-lever dual AC resonance tracking (iDART)—at AC drive amplitudes of 10 mV, 100 mV, and 1 V (annotated in units of the thermal-equivalent voltage, kBT/e). The oSF, oDART, and iDART signals contain contributions from electrostatic interactions, in-plane motion, and vertical piezoresponse, whereas iSF measurements are dominated by the vertical piezoresponse alone. At 10 mV (~0.4 kBT/e), oSF, oDART, and iSF are largely noise-limited with minimal domain contrast; in contrast, iDART preserves clear 180° phase reversal and measurable amplitude contrast. At 100 mV, domain structure becomes resolvable in most modes with the exception of oSF, while iDART continues to exhibit the highest contrast and most robust phase separation. At 1 V, oSF begins to reveal phase contrast consistent with both iDART and oDART. Together, these results demonstrate that iDART provides substantially improved signal-to-noise ratio, enabling functional piezoresponse imaging at millivolt-scale excitation biases. (b) Representative contact-resonance spectrum illustrating the dual-frequency tracking scheme. Two drive frequencies positioned above and below the contact resonance (separation Δf = 6 kHz) sample the resonance shoulders to generate the DART error signal. Curves (colored) show resonance fits with corresponding quality factors (Q) and resonance gains (G), illustrating how resonance enhancement combined with low-noise interferometric detection underpins the superior signal-to-noise performance of iDART.

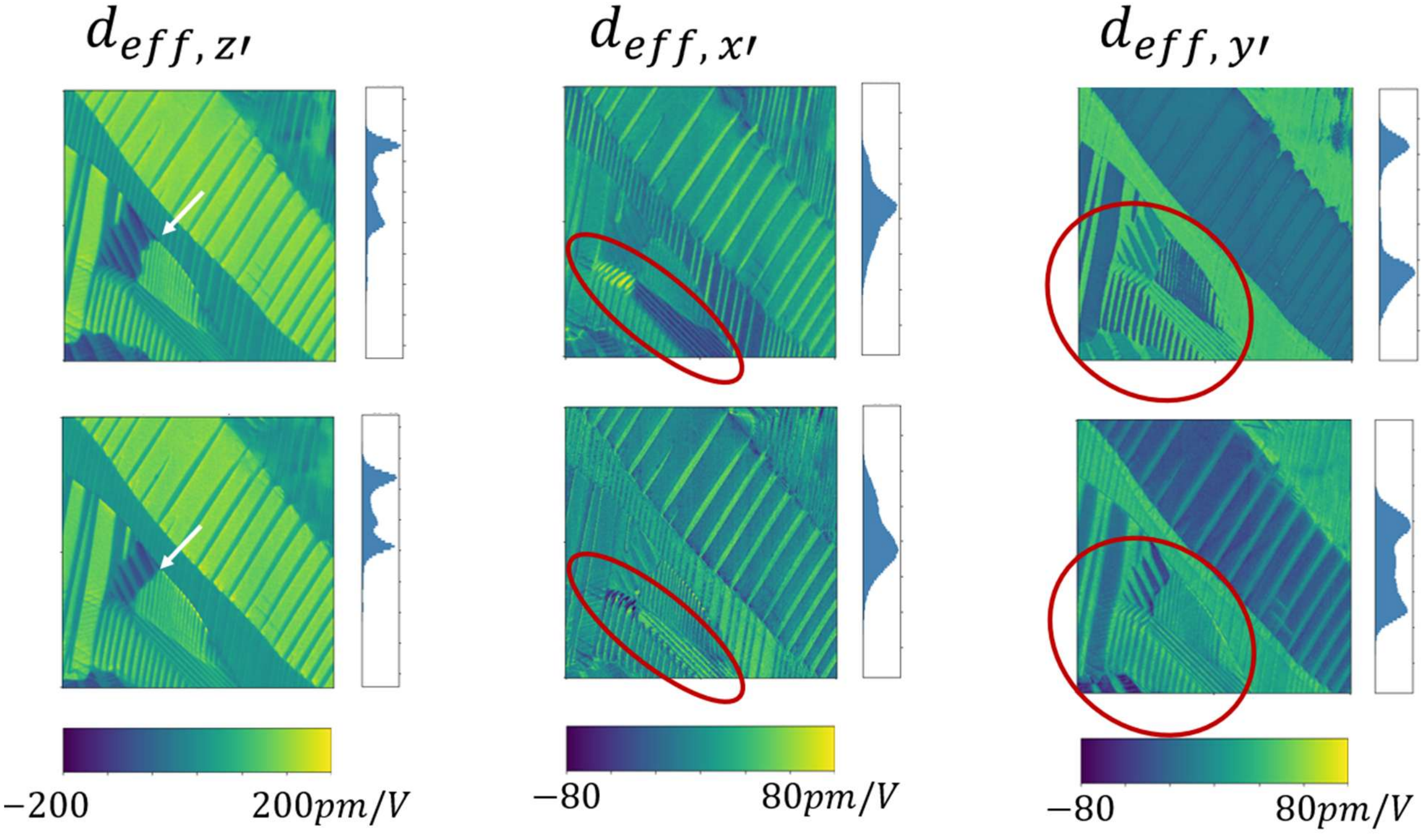

Figure S3. High AC bias (7Volts, 42kHz) driven changes in PZT. The arrows and circled regions point out domain structures that have changed under the influence of the large AC bias.

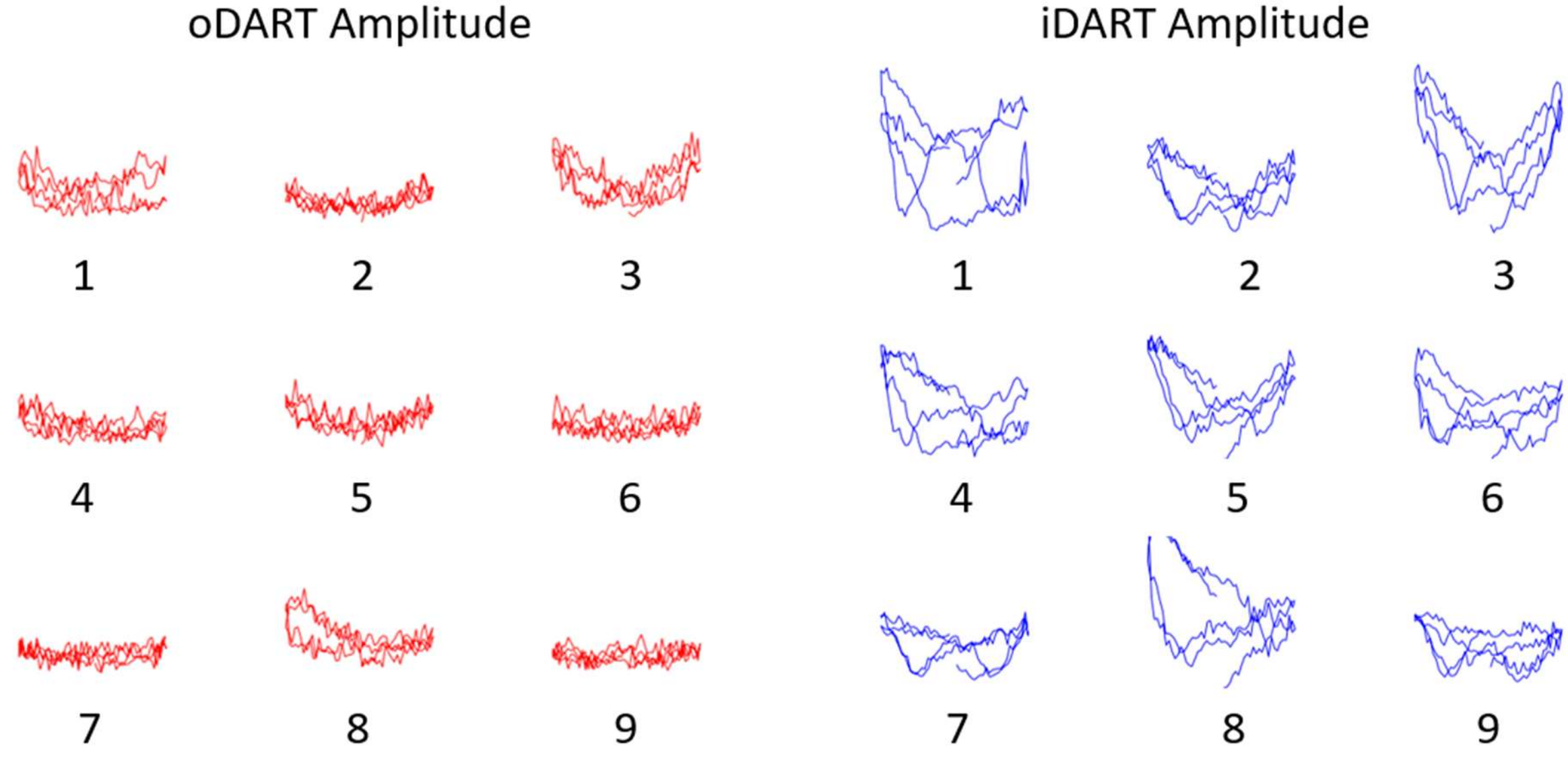

Figure S4. 3x3 array of position-dependent iSF and iDART SSPFM amplitude butterfly loops on the bare $Y{:}HfO_2$ films shown in Fig. 3.

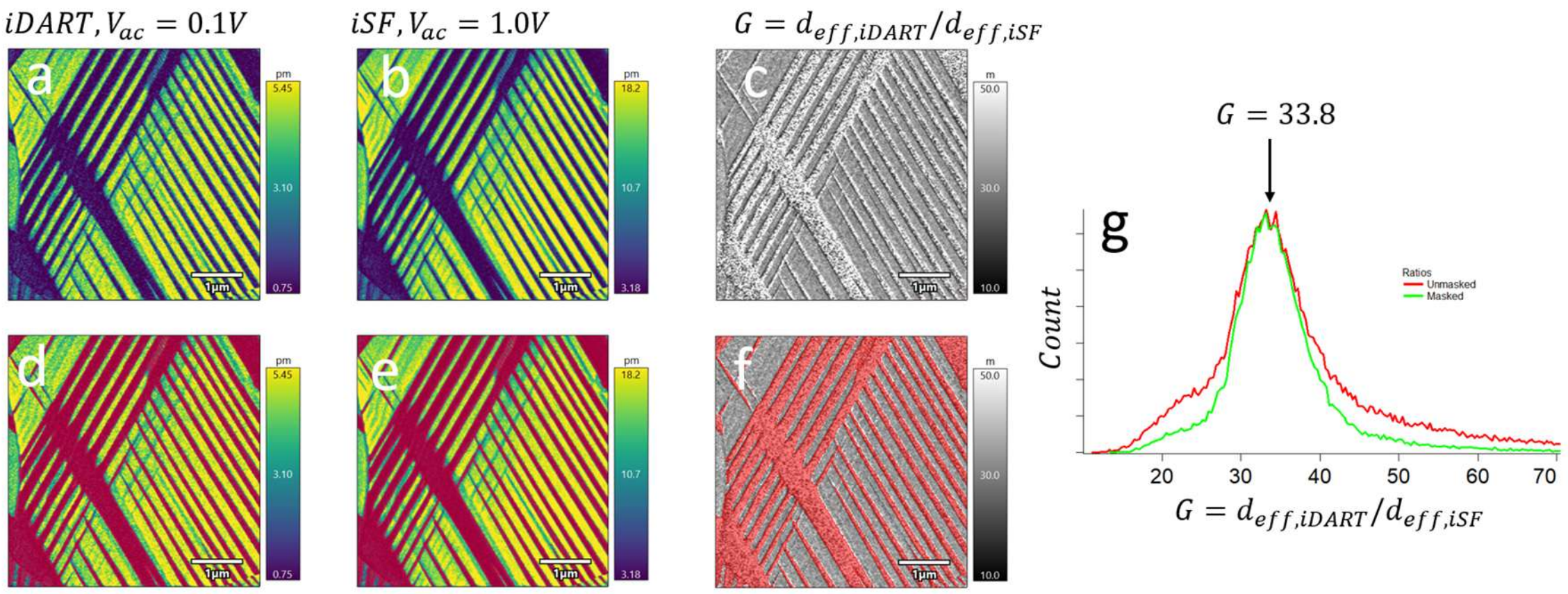

Figure S5. Ratio of iSF to iDART showing gain factor estimations for the entire images (a-c) and for images that were masked (d-f) to only contain the largest responses, omitting the small amplitude (and therefore noisier) data. In both cases, the peak ratio was $G \approx 34$. In general, we expect a distribution of gain values since the quality factor of the cantilever has substantial variations from pixel to pixel.

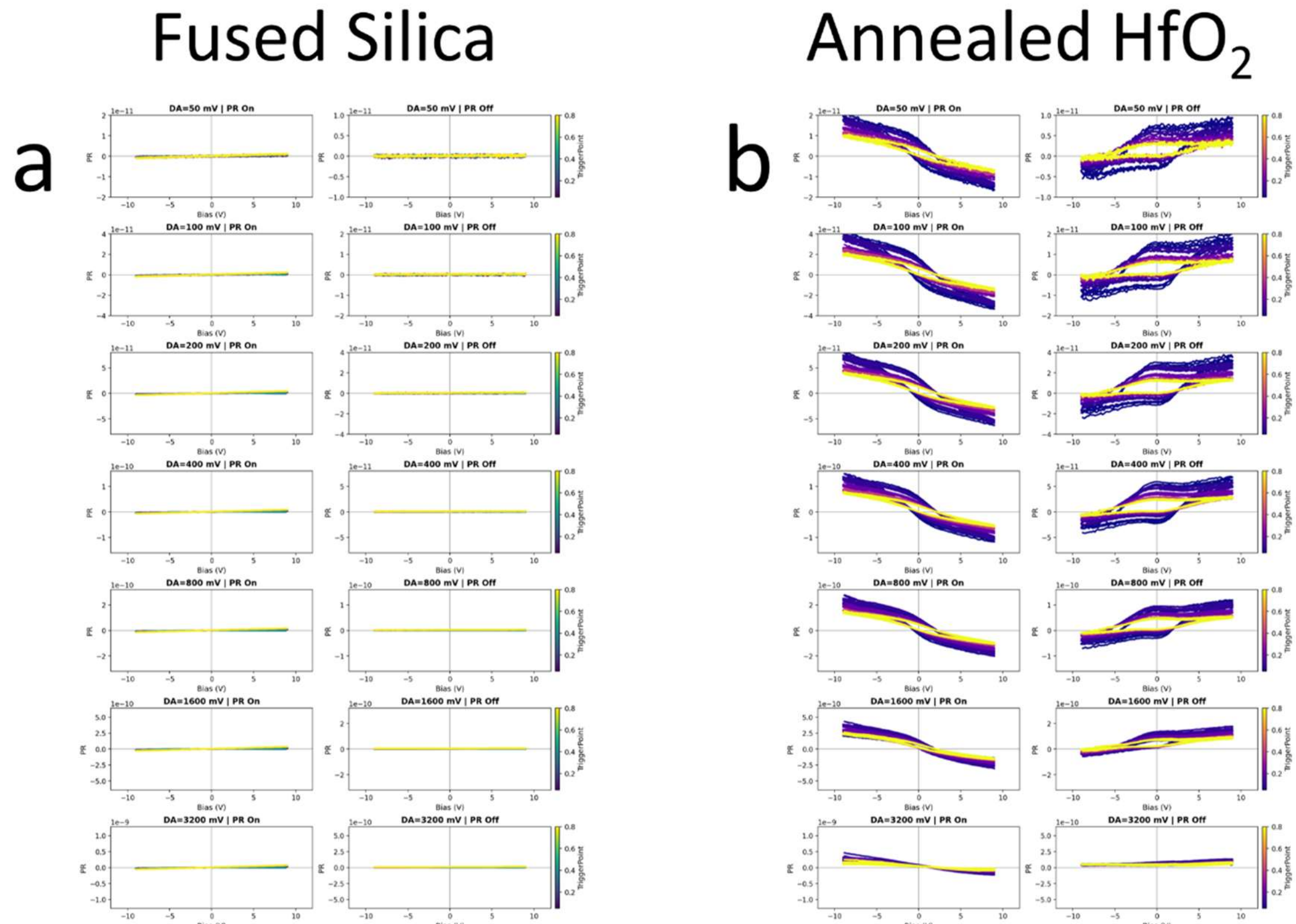

Figure S6. Averaged PR vs bias loops are shown for (a) fused silica, and (b) annealed (ferroelectric) $HfO_2$ film. In each panel, rows correspond to increasing drive amplitude (DA), while columns show PR-On (left) and PR-Off (right) responses. Within each subplot, curves are color-coded by trigger point. The z-axes are scaled for each subplot to emphasize loop shape and evolution with drive amplitude and trigger point. The loop area as a function of drive amplitude and trigger point (load) is summarized in Fig. 6 in the main text.

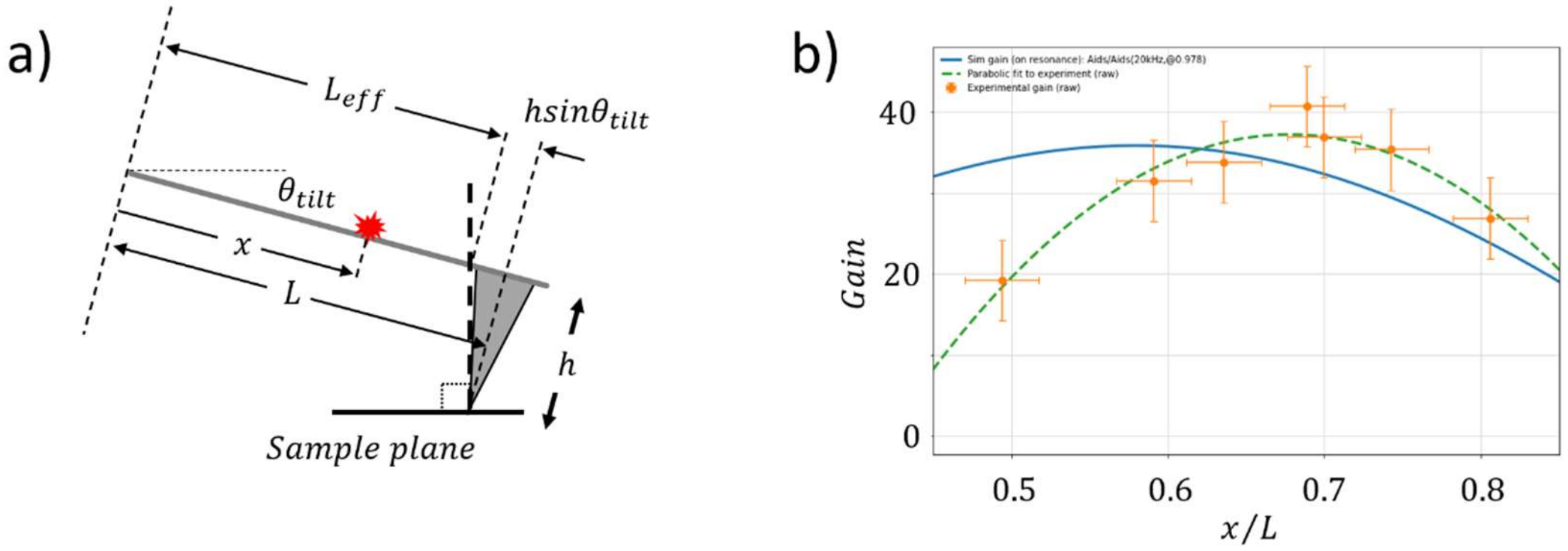

Figure S7. a Schematic illustrating the geometric origin of position-dependent resonant gain for a tilted cantilever. The cantilever of length $L$is inclined by an angle $\theta_{\text{tilt}}$relative to the sample plane, such that the effective interaction length $L_{\text{eff}}$and the local vertical lever arm $h\sin\ \theta_{\text{tilt}}$depend on the position $x$along the cantilever. The red star denotes the location of the interferometric detector spot. b Comparison of simulated and experimental effective resonant gain as a function of normalized position $x/L$. The solid blue curve shows the simulated on-resonance IDS gain, defined as the IDS amplitude at resonance divided by the IDS amplitude at $x/L \approx 0.98$for a 20 kHz (sub-resonant) drive. Orange symbols indicate experimentally measured gain with horizontal error bars of $\pm 5\mu m$ in $x$ and the vertical error bars . The dashed green curve is a parabolic fit to the experimental data, shown as a guide to the eye.

The normalization denominator is the cantilever response at 20 kHz (sub-resonance) evaluated at $x/L \approx 0.98$. We chose that value as the geometric null position for a cantilever oriented at 11 degrees with respect to the sample surface. $\frac{L_{eff}}{L} = \frac{(L-hsi\quad tilt)}{L} \approx 0.98$

$$\text{Gain}(x/L\,, f_{CR}) = \frac{A_{\text{int}}(x/L\,,\ f_{CR})}{A_{\text{int}}(x_{\text{ref}}/L = 0.98, f_{\text{ref}} = 20\,\text{kHz})}$$

Both the simulated on-resonance gain exhibits a broad, smoothly varying spatial dependence with a maximum in the mid-to-upper region of the cantilever, reflecting the combined effects of mode shape, electrostatic forcing distribution, and detection geometry. The experimental data show a similar overall trend, with a peak gain occurring at comparable values of $x/L$, and are well captured by a simple parabolic fit within the experimental uncertainties.

The quantitative differences between the simulated and fitted experimental curves likely arise from simplifications in the model, including validity of the Euler-Bernoulli model for real, non-uniform cantilevers, idealized boundary conditions and uniform material properties. In addition, most

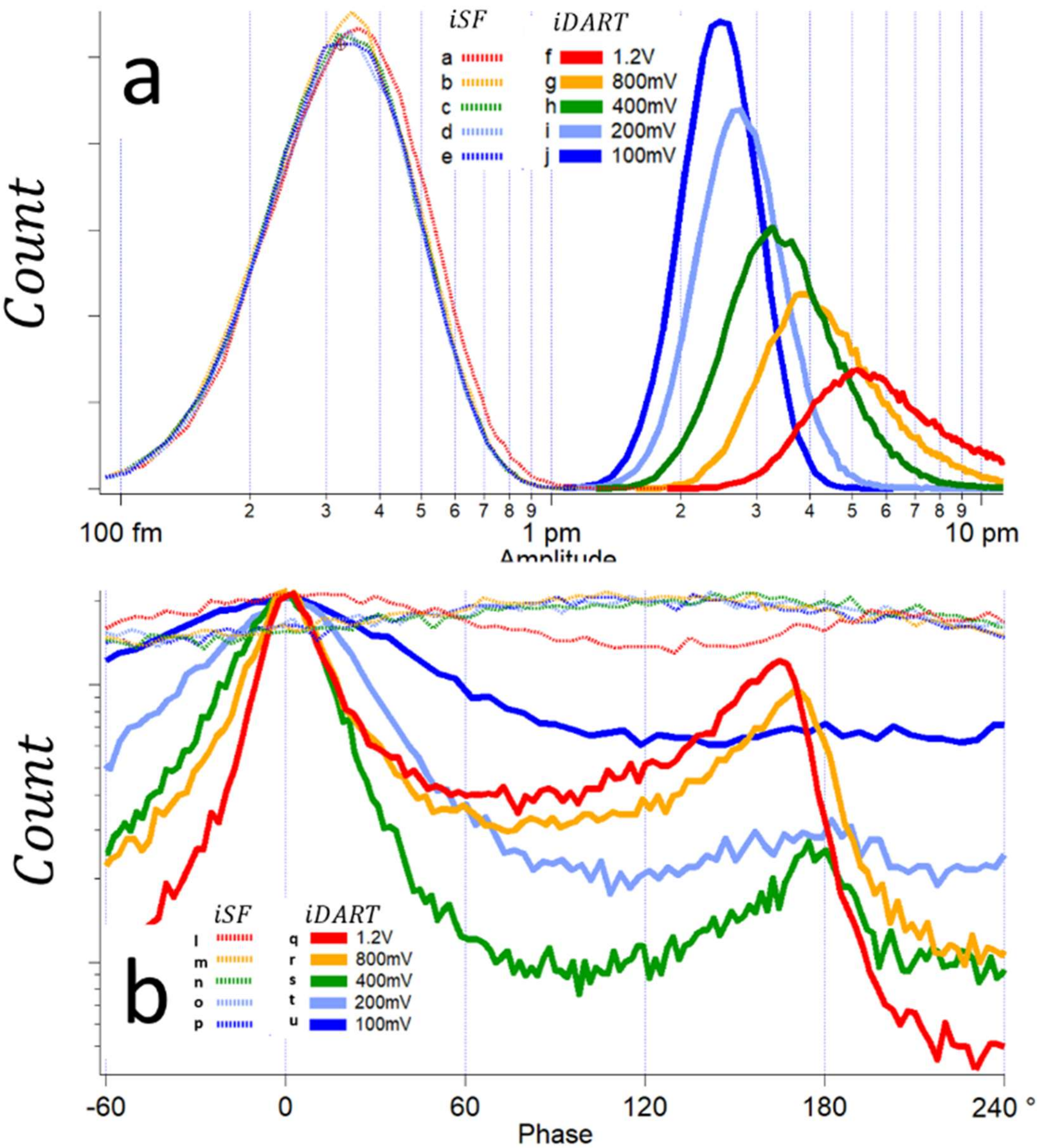

Figure S8. iSF (dashed) and iDART(solid) histograms for the amplitude (a) and phase (b) images of $Y:HfO_2$ films shown in Fig. 4 in the main text. Note that the iSF histograms remain nearly identical until Vac=1.2 V, when the iSF signal starts to emerge from the noise. Panel v shows the phase histograms. The iSF phase histograms (thin lines) show a very broad distribution with no clear peaks while the iDART histograms show clear bimodal peaks separated by ~180º degrees as expected for PFM of a ferroelectric sample.

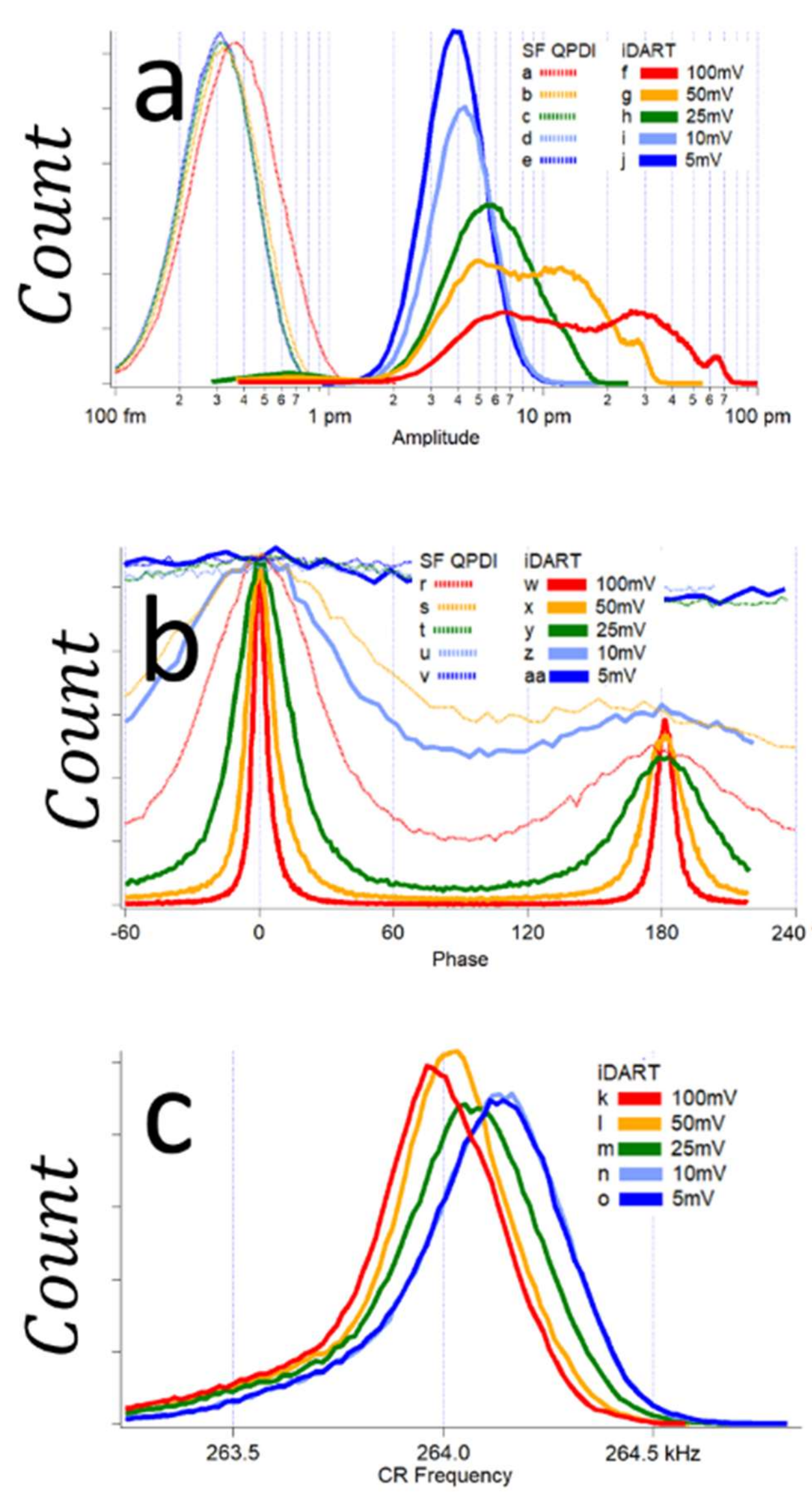

Figure S9. iSF (dashed) and iDART(solid) histograms for the amplitude (a), phase (b), and contact resonance frequency (c) images of the PZT films shown in Fig. 6 in the main text.

Background signals on three substrates:

These supplemental measurements demonstrate that substantial bias-dependent and loop-like PFM responses can be generated even on nominally non-piezoelectric reference substrates. Across fused Si, mica, and soda-lime glass, the apparent response depends strongly on both AC drive amplitude and readout condition, with clear differences between *PR*on and *PR*off. Fused Si generally shows the weakest response and smallest drive dependence, whereas mica and soda-lime glass exhibit larger apparent slopes, stronger applied load dependence, and more pronounced systematic trends with bias. The persistence of these responses in non-ferroelectric materials indicates that hysteresis-like PFM loops are not, by themselves, sufficient evidence for ferroelectric switching. Instead, they can arise from electrostatic forces, capacitive coupling, surface charging, ionic motion, adsorbate-mediated effects, or other bias-dependent interactions between the tip, cantilever, and sample surface. The scaling analysis further shows that these artifacts are not simply random noise: the normalized loop areas and fitted slopes vary systematically with drive amplitude and substrate, implying that the measured signal is influenced by material-dependent surface and electrostatic properties. These controls therefore reinforce the need to compare *PR*on and *PR*off responses, evaluate drive-amplitude dependence, and include non-ferroelectric reference materials when interpreting PFM hysteresis measurements.

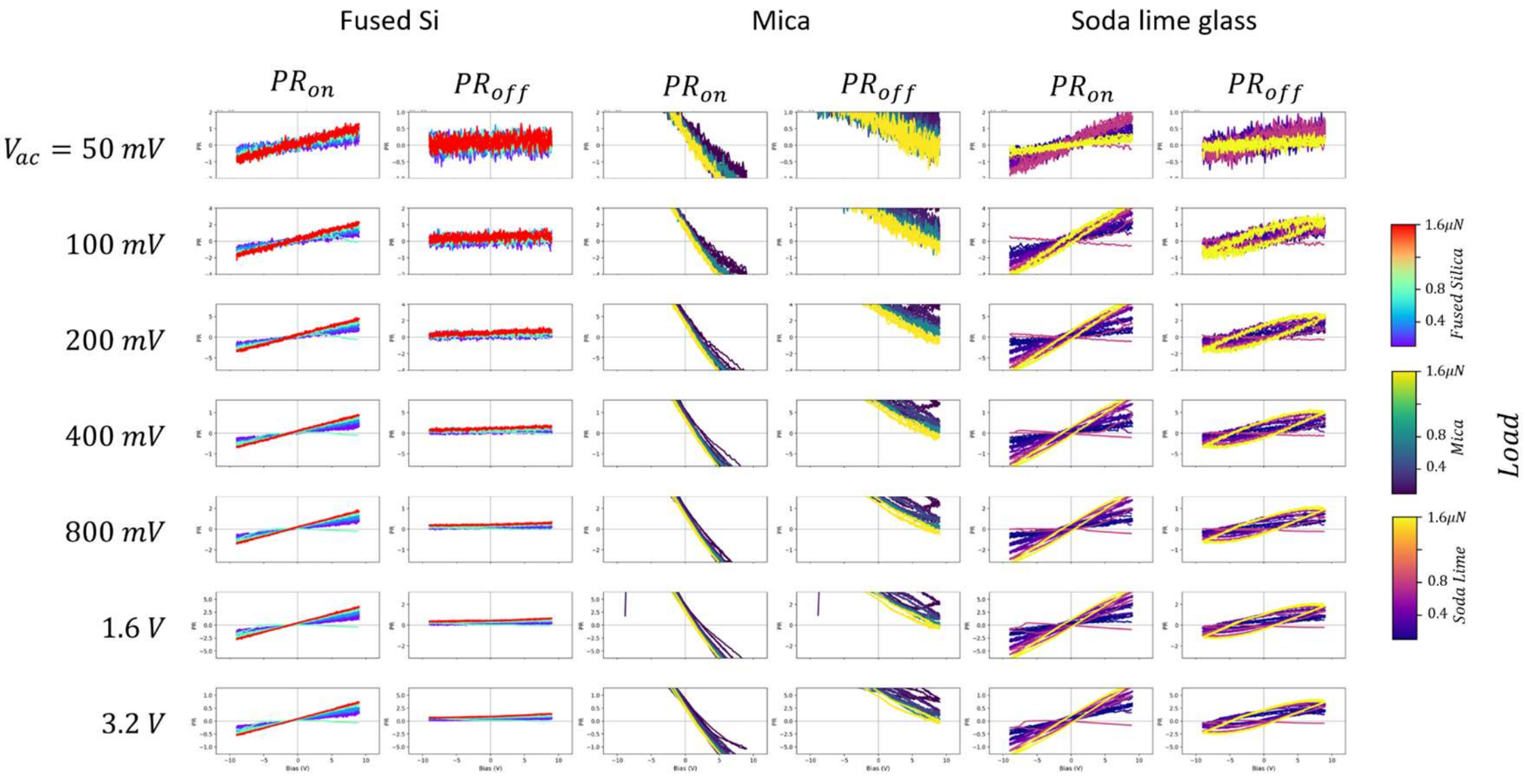

Figure S10. Bias-dependent apparent PFM response on non-piezoelectric reference substrates. Representative local hysteresis loops measured on fused Si, mica, and soda-lime glass as a function of AC drive amplitude, from 50 mV to 3.2 V. For each substrate, responses are shown for measurements acquired with the DC bias applied during readout ($PR_{on}$) and with the DC bias removed during readout ($PR_{off}$). Curves are colored by the applied load, with separate color scales for each material. Despite the absence of intrinsic ferroelectricity, mica and soda-lime glass substrates exhibit systematic bias-dependent slopes and loop-like responses. The persistence of these trends under different drive amplitudes and applied load conditions indicates that both bias-dependent responses, including hysteresis can arise from electrostatic, capacitive, localized chemistry or surface-mediated artifacts rather than true piezoelectric switching. Starting from the top, with $V_{ac} = 50\ mV$ to $V_{ac} = 3.2\ V$, the vertical axis range for $PR_{on}$ is $20, 40, 80, 160, 320, 640\ and\ 1280\ pm$ and for $PR_{off}$ is $10, 20, 40, 80, 160, 320\ and\ 640\ pm$. As discussed above, these are the as-measured amplitudes and have not been processed according to Gannepalli et al.[22]

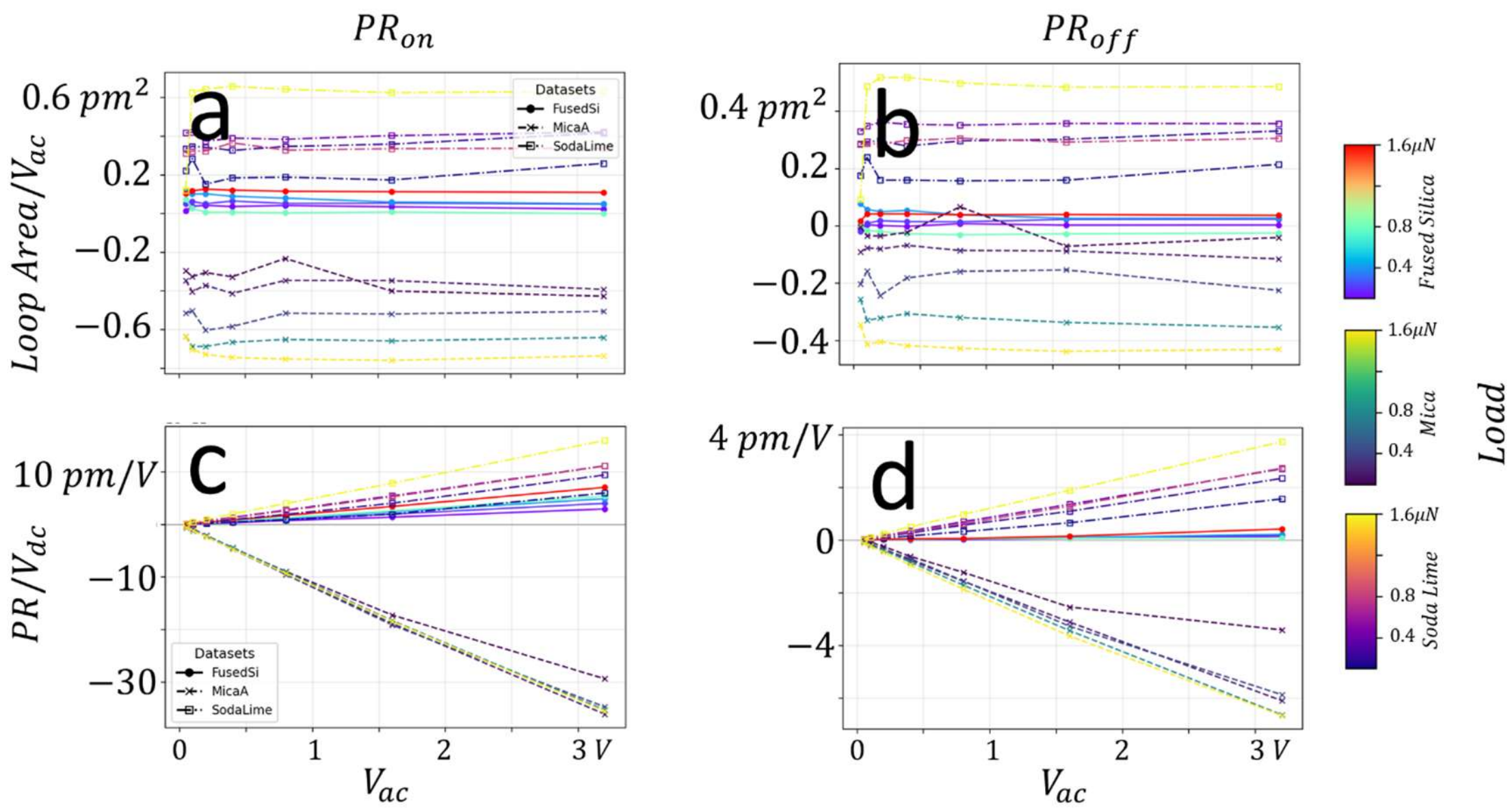

Figure S11. Drive-amplitude and load dependence of apparent PFM response on non-ferroelectric reference substrates. This Figure summarizes the local PFM spectroscopy measurements acquired on fused silica, mica, and soda-lime glass as a function of AC drive amplitude, $V_{ac}$, and applied normal load. Panels (a,b) show the hysteretic loop area normalized by AC drive amplitude, Loop Area/$V_{ac}$, for measurements acquired with the DC bias applied during readout, $PR_{on}$, and removed during readout, $PR_{off}$, respectively. Panels (c,d) show the corresponding normalized linear response, $PR/V_{ac}$, extracted from fits of the apparent piezoresponse versus DC bias for $PR_{on}$and $PR_{off}$, respectively. Marker style denotes the reference substrate: fused silica, mica, and soda-lime glass. Color indicates applied load, with darker to brighter colors corresponding to increasing load from approximately 0.4 to 1.6 μN. These measurements demonstrate that nominally non-ferroelectric materials can exhibit measurable bias-dependent and hysteresis-like PFM responses, with the magnitude and sign of the apparent response depending strongly on substrate, drive amplitude, load, and readout condition.

The data show that the apparent PFM response of non-ferroelectric substrates is not simply random noise, but contains systematic material- and load-dependent trends. Fused silica shows the smallest response overall. Its normalized loop area remains close to zero across the measured $V_{ac}$range, and the fitted $PR/V_{ac}$slopes are comparatively small in both $PR_{on}$and $PR_{off}$. This supports its use as the lowest-background reference substrate among the materials tested. In contrast, mica and soda-lime glass exhibit substantially larger apparent responses. Mica, in particular, shows large negative normalized $PR/V_{ac}$slopes that become increasingly pronounced with increasing $V_{ac}$, especially in $PR_{on}$. Soda-lime glass shows positive normalized slopes and comparatively large positive loop areas, also with a clear dependence on load and drive amplitude. The contrast between $PR_{on}$and $PR_{off}$further emphasizes that these signals are influenced by bias-dependent processes rather than intrinsic piezoelectricity. The $PR_{off}$responses are generally reduced relative to $PR_{on}$, but they do not vanish, indicating that remnant charge, surface chemistry, ionic redistribution, capacitive effects, or other slow relaxation processes can contribute to the measured signal even after the DC bias is removed.

In soda-lime glass, mobile $Na^+$ ions[47] and hydration-derived species such as $H^+/H_3O^+$, $OH^-$,[48] and silanol groups can accumulate or redistribute near the surface under applied bias, which may produce electrostatic or ionic contributions to the apparent PFM response. Similarly, freshly cleaved mica exposes a potassium-ion-rich basal surface, [49], [50] and subsequent exposure to ambient humidity may introduce hydrated or mobile ionic species such as $H^+/H_3O^+$ and $OH^-$,[51] which, in turn, may contribute to bias-dependent electrostatic or ionic artifacts. In fused silica, which lacks the abundant mobile alkali modifiers present in soda-lime glass and mica, bias-dependent surface artifacts are more likely to arise from hydroxylated silica surface chemistry, adsorbed water, and protonic charge redistribution. The surface of amorphous silica commonly contains silanol groups (Si–OH) and siloxane groups (Si–O–Si), and exposure to humidity produces interfacial water layers whose structure depends on the density and distribution of these sites. Water can also react with silica to form additional silanol groups, with proton or hydronium-mediated processes providing a possible pathway for charge redistribution under applied bias.[52], [53] These effects could contribute weak electrostatic or ionic components to an electromechanical response, even in the absence of intrinsic piezoelectricity.

These results reinforce the need for reference measurements and suggest that of the three substrates, Fused Si provides the superior piezoresponse null-reference.